\documentclass[pra,reprint,superscriptaddress,amsmath,amssymb,floatfix]{revtex4-2}

\usepackage{bm}
\usepackage{graphicx}
\usepackage{mathtools}
\usepackage{physics}
\usepackage{microtype}
\usepackage{xcolor}
\usepackage{hyperref}
\usepackage{orcidlink}
\hypersetup{colorlinks=true,linkcolor=blue,citecolor=blue,urlcolor=blue}
\newcommand{\ii}{\mathrm{i}}
\newcommand{\Om}{\Omega_g}

\usepackage{amsthm}
\usepackage{booktabs}
\usepackage{mathrsfs}
\renewcommand{\dd}{\mathrm{d}}

\newcommand{\eps}{\varepsilon}

\newcommand{\im}{\operatorname{Im}}
\newtheorem{theorem}{Theorem}[section]

\begin{document}

\title{Electrons in a hard-wall cylinder with helicoidal metric: time-reversal pairing, cubic anticrossings, and twist-activated dipole transitions}

\author{Humberto Belich Jr.\,\orcidlink{0000-0002-8795-1735}}
\affiliation{Departamento de F\'isica, Universidade Federal do Esp\'{\i}rito Santo, Vit\'oria, Esp\'{\i}rito Santo 29075-910, Brazil}
\author{Edilberto O. Silva\,\orcidlink{0000-0002-0297-5747}}
\email{edilberto.silva@ufma.br}
\affiliation{Programa de P\'os-Gradua\c{c}\~ao em F\'isica \& Coordena\c{c}\~ao do Curso de F\'{\i}sica -- Bacharelado, Universidade Federal do Maranh\~ao, S\~ao Lu\'{\i}s, Maranh\~ao 65080-805, Brazil}
\date{\today}

\begin{abstract}
We study the scalar Schr\"odinger equation in a helicoidal metric when the axial direction is closed by Dirichlet walls built into the operator domain. With hard walls the axial derivative cannot be replaced by a traveling-wave factor, and in a rotationally symmetric spinless cavity time reversal makes the levels with azimuthal numbers $m$ and $-m$ exactly degenerate. We write the Hamiltonian as a Hermitian coupled-subband matrix and give an exact gauge-adapted variational representation. For a finite radius the leading twist correction is quadratic and changes sign at a critical aspect ratio; for an unbounded radial domain it is linear in $|\Omega_g|$. Retuning the aspect ratio through an isolated radial--axial resonance within one $m$ sector opens an avoided crossing with gap $\Delta_{\mathrm{gap}}/E_R=1.648775\ldots|\Omega_gR^2/L|^3$ to leading order. A transverse dipole transition forbidden without twist acquires an oscillator strength quadratic in the twist, with leading coefficient $0.0559408010$ for $R/L=2$ from perturbation theory. Dipole sum rules and the convergence of the currents serve as checks on the numerics. We also give the inverse-mass tensor required for a scalar realization and re-examine each statement and figure of the earlier article [Opt. Quantum Electron. \textbf{58}, 113 (2026)] within the corrected model.
\end{abstract}
\keywords{Helicoidal geometry; Dirichlet confinement; orbital dynamics; subband mixing; geometric susceptibility; dipole sum rules}
\maketitle

\section{Introduction}\label{sec:intro}
Geometry enters wave mechanics in at least three ways: through the kinetic metric of the space itself, through the reduction of an ambient problem to a constrained surface or curve, and through an engineered material response. Each route leads to its own operator. The Laplace--Beltrami operator of a three-dimensional metric differs from the thin-layer geometric potential of an embedded surface \citep{daCosta1981}, and a scalar optical analogy does not by itself fix the Maxwell constitutive tensors \citep{Crudo2009}. The model studied here is of the first kind. The other two come up again when we discuss how it could be realized and how it relates to optics.

The classical and quantum action principles in curved spaces were reviewed by DeWitt \citep{DeWitt1957}. In condensed matter the question became concrete with the thin-layer procedure of Jensen and Koppe and of da Costa, in which a particle squeezed onto a surface or a curve by a steep confining potential acquires an effective potential that depends on the curvature of the embedding \citep{JensenKoppe1971,daCosta1981,daCosta1982}. Later work showed that the reduced Hamiltonian depends on how the constraint is imposed \citep{Kaplan1997}, extended the reduction to general embeddings and to external electric and magnetic fields \citep{SchusterJaffe2003,FerrariCuoghi2008}, and estimated the resulting energy shifts in nanostructures \citep{EncinosaEtemadi1998}. Geometry-driven changes of electronic properties have been predicted for deformed cylinders \citep{Shima2009} and observed experimentally \citep{Onoe2012}. Strained bilayers that roll up into tubes and helices \citep{Prinz2000,SchmidtEberl2001} made it possible to measure magnetotransport in curved two-dimensional electron gases \citep{Shaji2007,Vorobev2007}, and electrons in curved and twisted quantum wires, including spin--orbit coupling on curved surfaces, have been described by effective low-dimensional models \citep{EntinMagarill2001,Entin2002,MagarillEntin2003,MagarillChaplikEntin2005}.

A parallel line of work in mathematical physics concerns the Dirichlet Laplacian in tubes. Bending a tube of constant cross section produces bound states below the threshold of the essential spectrum \citep{ExnerSeba1989,GoldstoneJaffe1992,DuclosExner1995,Bulla1997}, and the torsion of the reference curve modifies them \citep{ClarkBracken1996}. Twisting a tube whose cross section is not rotationally symmetric acts in the opposite direction: it produces a Hardy inequality, so that small attractive perturbations no longer create bound states, and it accelerates the decay of the heat semigroup \citep{Ekholm2008,KrejcirikZuazua2010}. Periodic and locally perturbed twists have also been analyzed \citep{ExnerKovarik2005,Briet2009}, and Ref.~\citep{ExnerKovarik2015} reviews the field. In all these problems the ambient space is Euclidean, and the twist of a circular cross section can be removed by rotating the coordinates. The helicoidal metric considered here is different. Its scalar curvature is $-2\Om^2$, and the twist stays in the kinetic operator even for a circular cylinder.

The same metric appears in the geometric theory of defects, where dislocations and disclinations are encoded in the torsion and curvature of an effective three-dimensional geometry \citep{Kleinert1989,KatanaevVolovich1992,PuntigamSoleng1997,Moraes2000}. In this description a screw dislocation acts on electrons through a modified kinetic operator and a topological phase \citep{Bausch1998,Bausch1999,FurtadoBezerraMoraes2000}, and it shifts the Landau levels of a charged particle \citep{FurtadoMoraes1999,Marques2001}. The metric of Eq.~\eqref{eq:metric} has been associated with a uniform density of parallel screw dislocations; the axial momentum then acts as a charge in a uniform effective magnetic field, and the spectrum takes the form of elastic Landau levels \citep{SilvaNetto2008,SilvaNetto2026}. This links the model to the Landau and Fock--Darwin problems \citep{Landau1930,Fock1928,Darwin1931,Dunne1992} and to flux-induced phases \citep{AharonovBohm1959}. Darwin had already shown that the boundary of a sample enters the orbital magnetism of free electrons \citep{Darwin1931}, and the walls are also at the center of the present paper. Mathematically, Eq.~\eqref{eq:metric} is the left-invariant metric of the Heisenberg group, the Nil geometry among Thurston's eight model geometries \citep{Thurston1982,Scott1983}, and Laplacian spectra on its compact quotients have been computed \citep{GordonWilson1986,Andriot2018}.

Because the effective field is proportional to the axial momentum, the model is also related to synthetic gauge fields, in which an effective magnetic field is built into the kinetic operator by design. Laser-induced gauge potentials produce such fields for ultracold atoms \citep{Lin2009,Dalibard2011,Goldman2014}. Nonuniform strain in graphene generates pseudomagnetic fields \citep{Guinea2010,Vozmediano2010}, observed above $300$~T in nanobubbles \citep{Levy2010}, together with a position-dependent velocity tensor \citep{deJuan2012}. Strained photonic lattices and nonplanar optical resonators support photonic Landau levels \citep{Rechtsman2013,Schine2016}.

Geometry can also be imposed on light. Pleba\'nski showed that a gravitational field acts on electromagnetic waves like an anisotropic medium \citep{Plebanski1960}, a result that underlies transformation optics \citep{Leonhardt2006,Pendry2006,LeonhardtPhilbin2006} and its metric formulation \citep{Crudo2009}. Light confined to curved dielectric surfaces obeys wave equations in a curved two-dimensional space \citep{BatzPeschel2008}, and experiments have separated intrinsic from extrinsic curvature, measured intensity correlations, controlled light in curved nanophotonic structures, and observed accelerating wave packets \citep{Schultheiss2010,Schultheiss2016,Bekenstein2017,Patsyk2018}. Femtosecond-laser-written waveguides reproduce the geometric potential of a constrained quantum particle \citep{Szameit2010} together with many other quantum-optical analogies \citep{Longhi2009,SzameitNolte2010}. Helically twisted photonic crystal fibers couple the twist to the orbital angular momentum of light \citep{Allen1992}; they show orbital-angular-momentum resonances, optical activity and helical Bloch modes, and a twist alone can guide light in a coreless fiber \citep{Wong2012,Xi2013,Xi2014,Beravat2016,Russell2017}. For the screw-dislocation metric itself, minimally coupled Maxwell theory gives a Landau-type orbital propagation at leading paraxial order \citep{SilvaRevisiting2026}, and an additional axial electromagnetic response produces torsional optical activity \citep{SilvaSourceDriven2026}. Each of these analogies reproduces one particular operator; as noted above, a scalar wave equation in curved space does not fix the vector constitutive relations \citep{Plebanski1960,Crudo2009}.

A physical realization of the scalar model requires an effective-mass operator with a specific position-dependent inverse-mass tensor. The ordering of such operators and the matching conditions at heterointerfaces have a long history \citep{BenDanielDuke1966,MorrowBrownstein1984}, and envelope-function theory gives the conditions under which a single-band description holds \citep{Burt1992,Foreman1995,Pistol1999}. On the spectroscopic side, far-infrared absorption of laterally confined electrons resolves their discrete levels in a magnetic field \citep{SikorskiMerkt1989}, and Kohn's theorem shows that interactions do not shift the cyclotron frequency of a translationally invariant electron gas \citep{Kohn1961}. Levels of the same symmetry avoid each other when a parameter is varied \citep{vonNeumannWigner1929}, so avoided crossings and the transfer of oscillator strength between nearby lines show which symmetries a perturbation breaks.

Reference~\citep{SilvaOQE2026} combined an axial traveling wave with hard walls and obtained a spectrum with a term linear in both the azimuthal integer $m$ and the Dirichlet index $n$. With hard walls this substitution is not allowed, because the traveling wave does not belong to the domain of the axial derivative: the hard-wall eigenfunctions are standing waves, and the mixed derivative has no diagonal matrix element in a single real axial mode. Weak intersubband mixing does not rescue the result, because the symmetry of the full confined Hamiltonian forbids any $m\leftrightarrow-m$ splitting. Section~\ref{sec:correction} and Appendix~\ref{app:status} go through the statements of Ref.~\citep{SilvaOQE2026} one by one and re-derive those that have a counterpart in the corrected model.

Here we ask how longitudinal confinement changes the spectrum and the probability currents of the three-dimensional scalar Hamiltonian. We define the self-adjoint model, keep the traveling-wave solution where it applies, and obtain the finite-cylinder spectrum analytically where possible and numerically elsewhere, without replacing $\partial_z$ by $\ii k_n$. Apart from the checks of the published optical procedures in Sec.~\ref{sec:opt}, all numerical results come from the corrected Hamiltonian, and none are taken from Ref.~\citep{SilvaOQE2026}; the anticrossing and dipole coefficients were computed by direct Bessel quadrature and cross-checked in an independently assembled basis, for the stated truncations. We also ask whether the twist-induced subband mixing shows up in individual dipole oscillator strengths and at what order in the twist an avoided crossing opens, and we determine the kinetic tensor, boundary conditions and symmetries that a physical realization would have to reproduce.

Our new results concern the Dirichlet cylinder itself. The twist response changes sign with the aspect ratio, an avoided crossing opens at third order in the twist, and the twist activates dipole transitions that are forbidden without it. Earlier work on Landau-like levels for a uniform defect density \citep{SilvaNetto2008,SilvaNetto2026}, on the effective one-dimensional theory of twisted noncircular wires \citep{Entin2002}, and on Laplacian spectra of compact Heisenberg nilmanifolds \citep{Andriot2018} dealt with other domains and boundary conditions; what we add is the finite Dirichlet domain and its coupled-subband spectrum. We also give the local inverse-mass tensor that a divergence-form scalar operator must have to reproduce the model. This is a condition on the operator, and we make no claim that a particular semiconductor structure meets it. Programmable circuits and vector Maxwell propagation are left for separate work.

The paper is organized as follows. Sections~\ref{sec:model}--\ref{sec:weak} set up the model, the Dirichlet cavity and the weak-twist theory. Section~\ref{sec:spectra} gives spectra and currents, Sec.~\ref{sec:spectro} dipole spectroscopy, and Sec.~\ref{sec:realization} the conditions for a physical realization. Section~\ref{sec:disc} discusses the optical model and Ref.~\citep{SilvaOQE2026}, and Sec.~\ref{sec:concl} concludes; Appendix~\ref{app:status} lists the status of each statement of Ref.~\citep{SilvaOQE2026}.

\section{Helicoidal metric, scalar Hamiltonian, and traveling waves}\label{sec:model}
We first fix the metric, the scalar Hamiltonian, and the domain on which it acts (Sec.~\ref{sec:geom}). On an open axis the axial wave number is conserved and the twist produces a Landau problem (Sec.~\ref{sec:travel}). The cavity of Sec.~\ref{sec:cavity} differs from this case only through its axial walls.

\subsection{Metric, Hamiltonian, and operator domain}\label{sec:geom}
We use cylindrical coordinates $r\ge0$, $0\le\phi<2\pi$, $z\in\mathbb R$ and the metric \citep{KatanaevVolovich1992,SilvaNetto2008,Scott1983,GordonWilson1986,SilvaRevisiting2026,SilvaSourceDriven2026}
\begin{equation}\label{eq:metric}
\dd s^2=\dd r^2+r^2\dd\phi^2+(\dd z+\Om r^2\dd\phi)^2,
\qquad [\Om]=\mathrm{length}^{-1}.
\end{equation}
For constant real $\Om$ the metric, its inverse, and its volume element are
\begin{equation}
\begin{aligned}
(g_{ij})&=\begin{pmatrix}1&0&0\\0&r^2+\Om^2r^4&\Om r^2\\0&\Om r^2&1\end{pmatrix},\\
(g^{ij})&=\begin{pmatrix}1&0&0\\0&r^{-2}&-\Om\\0&-\Om&1+\Om^2r^2\end{pmatrix},\\
\sqrt g&=r.
\end{aligned}
\end{equation}
An orthonormal coframe and its dual are
\begin{align}
&\vartheta^{\hat r}=\dd r,\quad \vartheta^{\hat\phi}=r\dd\phi,
\quad \vartheta^{\hat z}=\dd z+\Om r^2\dd\phi,\label{eq:coframe}\\
&e_{\hat r}=\partial_r,\quad e_{\hat\phi}=r^{-1}\partial_\phi-\Om r\partial_z,
\quad e_{\hat z}=\partial_z.\label{eq:frame}
\end{align}
The metric is the left-invariant metric of the Nil geometry \citep{Scott1983}, with scalar curvature $\mathcal R=-2\Om^2$. Because the Levi--Civita connection is torsion-free, we call $\Om$ a metric twist and not a Cartan torsion. In the geometric theory of defects the same metric is paired with a flat, metric-compatible defect connection whose torsion is $T^{\hat z}=2\Om\,\vartheta^{\hat r}\wedge\vartheta^{\hat\phi}$, and the Burgers flux gives $\Om=b\sigma/2$ for a uniform areal density $\sigma$ of parallel screw dislocations with Burgers vector $b$ \citep{SilvaNetto2008,SilvaRevisiting2026}. The scalar Hamiltonian below uses only the metric and its Levi--Civita connection, so this torsion does not enter it. Relating $\Om$ to the defect density of an actual elastic medium requires a separate identification of material parameters.

\begin{figure}[t]
\centering\includegraphics[width=\linewidth]{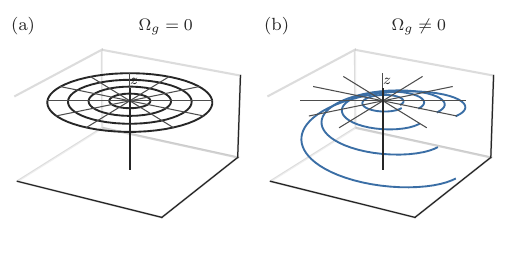}
\caption{Polar grid at $z=0$ (a) without twist and (b) with twist, where the circles are replaced by the curves orthogonal to $\partial_z$, $\dd z=-\Om r^2\dd\phi$, drawn for $r/R=0.25$, $0.5$, $0.75$, $1$ and $\Om R=0.1$. Radial lines remain orthogonal to $\partial_z$ at constant $z$; the curves through the circles rise by $2\pi|\Om|r^2$ per turn and do not close.}\label{fig:geometry}
\end{figure}

Figure~\ref{fig:geometry} shows what the cross term means geometrically. Since $e_{\hat z}=\partial_z$ is a unit vector, a curve is orthogonal to the axial Killing field $\partial_z$ when $\vartheta^{\hat z}$ vanishes along it, i.e., when $\dd z=-\Om r^2\dd\phi$. Radial lines at constant $z$ satisfy this condition, but a circle of constant $r$ and $z$ does not: the orthogonal curve through one of its points rises by $2\pi|\Om|r^2$ per turn and does not close. At $r=R$ and for the parameters of Eq.~\eqref{eq:params} the rise is $1.26\,\mu\mathrm{m}$, or $12.6L$. The description used in Ref.~\citep{SilvaOQE2026}, in which each transverse plane is rotated by $\Om\Delta z$, corresponds instead to $\dd r^2+r^2(\dd\phi+\Om\dd z)^2+\dd z^2$, which is flat space in helicoidal coordinates.

The scalar Hamiltonian for a constant effective mass $M>0$ is the Laplace--Beltrami operator of this metric \citep{DeWitt1957,FurtadoMoraes1999,SilvaNetto2008},
\begin{equation}\label{eq:H}
\begin{aligned}
H&=-\frac{\hbar^2}{2M}\Delta_g,\\
\Delta_g&=\frac1r\partial_r(r\partial_r)+\frac1{r^2}\partial_\phi^2\\
&\quad +(1+\Om^2r^2)\partial_z^2
 -2\Om\partial_\phi\partial_z.
\end{aligned}
\end{equation}
The quadratic form reads
\begin{equation}\label{eq:form}
\begin{aligned}
\langle\Psi,H\Psi\rangle
&=\frac{\hbar^2}{2M}\int r\,\dd r\,\dd\phi\,\dd z
\bigl(|\partial_r\Psi|^2\\
&\quad+|r^{-1}\partial_\phi\Psi-\Om r\partial_z\Psi|^2
+|\partial_z\Psi|^2\bigr).
\end{aligned}
\end{equation}
Equation~\eqref{eq:H} contains no da Costa-type potential, since no thin-layer reduction is made \citep{JensenKoppe1971,daCosta1981}. Its spectrum is defined only after a domain and boundary conditions are specified. We impose Dirichlet conditions on the finite cylinder
\begin{equation}\label{eq:domain}
\begin{aligned}
\mathcal D_{R,L}&=\{0<r<R,\ -L/2<z<L/2,\\
&\qquad\phi\in\mathbb S^1\},\\
\Psi|_{r=R}&=\Psi|_{z=\pm L/2}=0.
\end{aligned}
\end{equation}
Regularity at $r=0$ is assumed. In the sector $m=0$ the radial equation admits the local behaviors $F(r)\sim 1$ and $F(r)\sim\log r$ near the axis; the logarithmic branch has divergent radial Dirichlet energy $\int_0^\epsilon r|F'(r)|^2\,\dd r$ and is excluded by the Friedrichs form domain. For $|m|\ge1$ the regular behavior is $F\sim r^{|m|}$. No point interaction or extension parameter is placed on the axis, and the positive closed form of Eq.~\eqref{eq:form} defines a self-adjoint Friedrichs realization. We also consider $R\to\infty$ and a periodic axial direction. These are different operators, and their spectra cannot be obtained from one common separated solution.

\subsection{Traveling waves and the Landau-like sector}\label{sec:travel}
On the translationally invariant axis, $\Psi=e^{\ii m\phi+\ii kz}F(r)$, $m\in\mathbb Z$, is an admissible generalized eigenfunction. With $\beta=|\Om k|$ and $k\Om\neq0$, its radial operator is
\begin{equation}\label{eq:rad-trav}
\begin{aligned}
H_{m,k}&=-\frac{\hbar^2}{2M}
\left(\partial_r^2+r^{-1}\partial_r-m^2r^{-2}\right)\\
&\quad+\frac{\hbar^2k^2}{2M}
+\frac{\hbar^2\Om^2k^2r^2}{2M}
-\frac{\hbar^2\Om mk}{M}.
\end{aligned}
\end{equation}
The $r\in[0,\infty)$ solutions are labeled by the radial Laguerre quantum number $n_r=0,1,\ldots$; the azimuthal quantum number is $m\in\mathbb Z$. For axial Dirichlet modes we reserve $n_z=1,2,\ldots$; the letters $p,n$ below are axial-basis indices. As in the Fock--Darwin and Landau problems \citep{Fock1928,Darwin1931,Landau1930,SilvaNetto2008}, the solutions are
\begin{align}
F_{n_r,m,k}(r)
&=\mathcal{N}_{n_r,m}\,r^{|m|}e^{-\beta r^2/2}
L_{n_r}^{|m|}(\beta r^2),\nonumber\\
\mathcal{N}_{n_r,m}
&=\left[\frac{2\beta^{|m|+1}n_r!}
{\Gamma(n_r+|m|+1)}\right]^{1/2},\label{eq:laguerre}\\
E_{n_r,m}(k)&=\frac{\hbar^2k^2}{2M}
+\frac{\hbar^2|\Om k|}{M}(2n_r+|m|+1)\nonumber\\
&\quad-\frac{\hbar^2\Om mk}{M}.\label{eq:travel-spectrum}
\end{align}
We normalize $F$ by $\int_0^\infty r\,\dd r\,|F|^2=1$. Every term in Eq.~\eqref{eq:travel-spectrum} has the dimension of energy; in particular, the linear cross term carries the factor $\hbar^2/M$. At fixed $k$ the energies of $+m$ and $-m$ differ, but time reversal requires
\begin{equation}\label{eq:travTR}
E_{n_r,m}(k)=E_{n_r,-m}(-k).
\end{equation}
For a periodic axial direction both signs of $k$ belong to the spectrum. The guided electromagnetic modes of the same geometry have the same structure: their splitting between $+m$ and $-m$ at a fixed propagation direction is invariant under $(m,k_z)\to(-m,-k_z)$ \citep{SilvaSourceDriven2026}. At $k=0$ there is no twist-induced radial oscillator on the infinite disk. The family \eqref{eq:travel-spectrum} does not satisfy the boundary conditions \eqref{eq:domain}.

For fixed $k$, Eq.~\eqref{eq:rad-trav} is the symmetric-gauge Landau Hamiltonian of a charge in a uniform field with $qB_{\rm eff}=2\hbar\Om k$, shifted by $\hbar^2k^2/2M$ \citep{Landau1930,Fock1928,Dunne1992}. The cyclotron energy is $\hbar\omega_c=2\hbar^2|\Om k|/M$, and Eq.~\eqref{eq:travel-spectrum} becomes $E=\hbar\omega_c(N+\tfrac12)+\hbar^2k^2/2M$ with Landau index $N=n_r+[|m|-m\,\mathrm{sgn}(\Om k)]/2$. The term linear in $m$ is the paramagnetic coupling $-(qB_{\rm eff}/2M)L_z$, and the $1.0408$ meV splitting found in Ref.~\citep{SilvaOQE2026} at $k=\pi/L$ is $\hbar\omega_c$. The same mapping fixes the sense of the probability current. With the physical azimuthal current $J_{\hat\phi}=(\hbar/M)\im[\Psi^*(r^{-1}\partial_\phi-\Om r\partial_z)\Psi]$ of Sec.~\ref{sec:currents} and $\langle r^2\rangle=(2n_r+|m|+1)/|\Om k|$ for the states \eqref{eq:laguerre}, a planar eigenstate $\Psi=e^{\ii m\phi+\ii kz}F/\sqrt{2\pi}$ has
\begin{equation}\label{eq:Lkin}
\frac{M}{\hbar}\int\dd^2r\,r\,J_{\hat\phi}=m-\Om k\langle r^2\rangle=-(2N+1)\,\mathrm{sgn}(\Om k).
\end{equation}
The integrated circulation of every traveling eigenstate has the sign of $-\Om k$, whatever the sign of $m$.

For the value $k=\pi/L$ used in Ref.~\citep{SilvaOQE2026} and the parameters of Eq.~\eqref{eq:params}, $\omega\equiv\hbar\Om k/M=7.906\times10^{11}\,\mathrm{s}^{-1}$ and $\ell_{\rm ho}=(\Om k)^{-1/2}=79.8$ nm. The radial equation for $u=\sqrt r\,F$ has the effective potential
\begin{equation}\label{eq:Veffu}
\begin{aligned}
V_{\rm eff}(r)={}&\frac{\hbar^2(m^2-\frac14)}{2Mr^2}+\frac12M\omega^2r^2\\
&-\hbar\omega m+\frac{\hbar^2k^2}{2M},
\end{aligned}
\end{equation}
shown in Fig.~\ref{fig:travelprofiles}(a) for $m=1$. Its minimum lies at $r_0=(m^2-\frac14)^{1/4}\ell_{\rm ho}=74.3$ nm, with $V_{\min}=\hbar\omega(\sqrt{m^2-1/4}-m)+\hbar^2k^2/2M=1.565$ meV. These are the values of Fig.~2 of Ref.~\citep{SilvaOQE2026}; they hold for the reduced function $u$ of a traveling state, not for a level of the closed cavity. With the transverse normalization $\int\dd^2r\,|\psi|^2=1$, the $n_r=0$ states have $|\psi_{0,0}(0)|^2=1/(\pi\ell_{\rm ho}^2)$, and for $m=1$ and $\Om k>0$ the azimuthal component of Eq.~\eqref{eq:Jframe} is
\begin{equation}\label{eq:Jtravel}
\begin{aligned}
J_{\hat\phi}(r)&=\frac{\hbar}{M}|\psi_{0,1}|^2\Bigl(\frac{m}{r}-\Om kr\Bigr)\\
&=\frac{\hbar}{\pi M\ell_{\rm ho}^3}\,x(1-x^2)e^{-x^2},\qquad x=\frac{r}{\ell_{\rm ho}}.
\end{aligned}
\end{equation}
The density peaks at $r=\ell_{\rm ho}$, and there the local current changes sign. The current has a positive maximum at $0.468\ell_{\rm ho}=37.4$ nm and a negative minimum at $1.510\ell_{\rm ho}=120.5$ nm [Fig.~\ref{fig:travelprofiles}(b),(c)], and its first moment gives Eq.~\eqref{eq:Lkin}. The flat-space term $m/r$ alone, used in Eq.~(21) of Ref.~\citep{SilvaOQE2026}, gives a current of one sign with its maximum at $\ell_{\rm ho}/\sqrt2=56.4$ nm; the metric term reverses the circulation outside $r=\ell_{\rm ho}$.
\begin{figure}[t]
\centering\includegraphics[width=\linewidth]{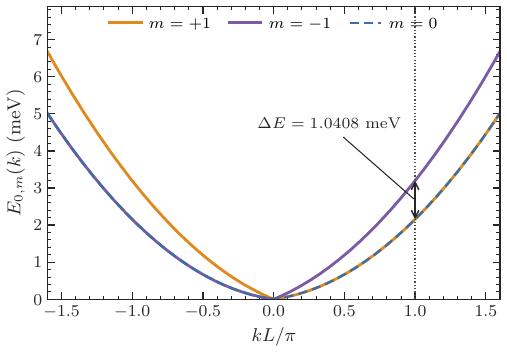}
\caption{Traveling-wave spectrum for an unbounded radial domain: the $n_r=0$ branches with $m=0,\pm1$ as functions of the axial wave number $k$, for $\Om>0$ and the material parameters of Eq.~\eqref{eq:params}. The dashed $m=0$ branch coincides with $m=+1$ for $k>0$ and with $m=-1$ for $k<0$ (Landau index $N=0$). At $k=\pi/L$ the two $|m|=1$ energies differ by $1.040818$ meV, and the full spectrum pairs $(m,k)$ with $(-m,-k)$. These states do not satisfy the Dirichlet conditions of the closed cavity.}\label{fig:travel}
\end{figure}
\begin{figure*}[t]
\centering\includegraphics[width=\textwidth]{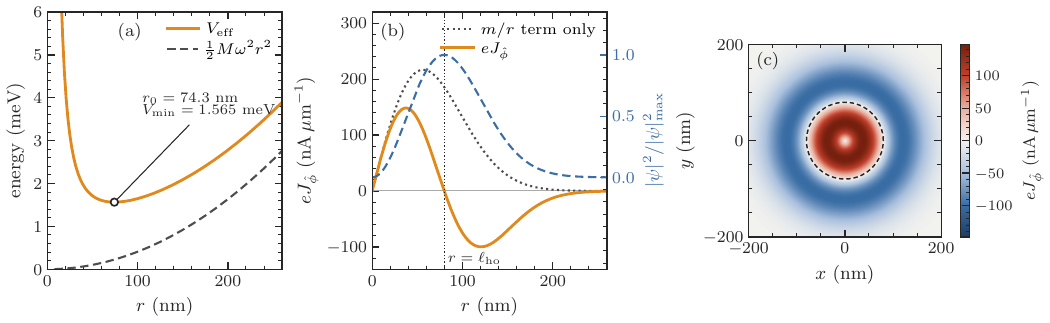}
\caption{Traveling state with $k=\pi/L$, $n_r=0$, $m=+1$ and the parameters of Eq.~\eqref{eq:params}. (a) Effective potential \eqref{eq:Veffu} of the reduced function $u=\sqrt rF$ (solid) and its harmonic part (dashed); the circle marks $r_0=74.3$ nm, $V_{\min}=1.565$ meV. (b) Azimuthal current $eJ_{\hat\phi}$ of Eq.~\eqref{eq:Jtravel} for one electron with transverse normalization (solid), the flat-space $m/r$ term alone (dotted), and the normalized density (dashed, right axis); the vertical line marks $r=\ell_{\rm ho}=79.8$ nm. (c) Map of $eJ_{\hat\phi}$; the dashed circle is $r=\ell_{\rm ho}$, where the current changes sign.}\label{fig:travelprofiles}
\end{figure*}

\section{The Dirichlet cavity}\label{sec:cavity}
Hard walls at the ends of the axis change the problem, because the axial derivative is no longer diagonal in the Dirichlet modes. We show first that the levels of $m$ and $-m$ coincide (Sec.~\ref{sec:dir}), then write the Hamiltonian as an exact matrix in a Bessel--sine basis (Sec.~\ref{sec:matrix}), and finally solve the $m=0$ sector exactly (Sec.~\ref{sec:m0}).

\subsection{Axial confinement and time-reversal symmetry}\label{sec:dir}
Axial confinement preserves rotational invariance. Writing
\begin{equation}\label{eq:reduce}
\Psi(r,\phi,z)=\frac{e^{\ii m\phi}}{\sqrt{2\pi}}u_m(r,z)
\end{equation}
reduces the Hamiltonian exactly to
\begin{equation}\label{eq:Hm}
\begin{aligned}
H_m=\frac{\hbar^2}{2M}\biggl[
&-\frac1r\partial_r(r\partial_r)+\frac{m^2}{r^2}\\
&-(1+\Om^2r^2)\partial_z^2
+2\ii m\Om\partial_z\biggr].
\end{aligned}
\end{equation}
Let $\mathcal K$ denote complex conjugation of $u_m$. Since the coefficients of the full Laplace--Beltrami operator are real,
\begin{equation}
\begin{gathered}
\mathcal K H_m(\Om)\mathcal K^{-1}=H_{-m}(\Om),\\
H_m(\Om)\text{ and }H_{-m}(\Om)
\text{ are isospectral}.
\end{gathered}
\end{equation}
For a spinless particle this is the action of time reversal on $e^{\ii m\phi}u_m$.
\begin{theorem}[Exact Dirichlet orbital pairing]\label{thm:TR}
If the scalar operator \eqref{eq:H} has real coefficients and its boundary/domain conditions are invariant under spinless time reversal and rotations about $z$, every eigenvalue in the $m$ sector also belongs to the $-m$ sector, with the same multiplicity. In particular, for the Dirichlet cylinder \eqref{eq:domain}, spinless time reversal \citep{Wigner1932} gives
\begin{equation}\label{eq:pair}
E_{\alpha,m}(\Om)=E_{\alpha,-m}(\Om),\qquad m\in\mathbb Z,
\end{equation}
where $\alpha$ orders levels within either isospectral sector.
\end{theorem}
\begin{proof}
If $H\Psi=E\Psi$ and $-\ii\partial_\phi\Psi=m\Psi$, then $H\Psi^*=E\Psi^*$, $-\ii\partial_\phi\Psi^*=-m\Psi^*$, and the real Dirichlet trace still vanishes. Complex conjugation is an invertible antiunitary map between the two sectors.
\end{proof}
This is not a Kramers degeneracy, since $\mathcal T^2=+1$; the two partners have different angular eigenvalues and are orthogonal. A boundary that breaks rotational symmetry can mix them into real standing waves, and time reversal alone then no longer forces a twofold degeneracy.

For constant $\Om$ the reflection $\mathcal P_z:z\mapsto-z$ about the center of the box gives in addition $E_{\alpha,m}(\Om)=E_{\alpha,m}(-\Om)$. At finite $R$ an isolated level is therefore analytic and even in $\Om$ near zero; the infinite radial domain is a singular limit (Sec.~\ref{sec:crossover}).

\subsection{Exact subband matrix and selection rules}\label{sec:matrix}
Choose the Dirichlet basis
\begin{equation}\label{eq:axbasis}
\begin{gathered}
\chi_n(z)=\sqrt{\frac2L}\sin\left[\frac{n\pi}{L}(z+L/2)\right],\\
n=1,2,\ldots,\qquad k_n=\frac{n\pi}{L}.
\end{gathered}
\end{equation}
Its derivative matrix is
\begin{equation}\label{eq:D}
\langle\chi_p|\partial_z|\chi_n\rangle=\frac1L D_{pn},\quad
D_{pn}=\begin{cases}\displaystyle\frac{4pn}{p^2-n^2},&p+n\ \text{odd},\\[5pt]0,&p+n\ \text{even}.\end{cases}
\end{equation}
Thus $D_{pp}=0$ and $D_{pn}=-D_{np}$: the derivative has no diagonal elements in the hard-wall modes and connects only opposite axial parities. On the Dirichlet domain the operator $-\ii\hbar\partial_z$ is symmetric but not self-adjoint, and it has no eigenvalues $\pm\hbar k_n$, although $-\partial_z^2$ has eigenvalues $k_n^2$.

For a finite radial boundary, use $\ell=|m|$ and the normalized Bessel basis
\begin{equation}\label{eq:Bessel}
\begin{gathered}
B_{\ell s}(r)=\frac{\sqrt2}{R|J_{\ell+1}(j_{\ell s})|}
 J_\ell(j_{\ell s}r/R),\\
\int_0^R r\,\dd r\,B_{\ell s}B_{\ell t}=\delta_{st}.
\end{gathered}
\end{equation}
where $j_{\ell s}$ is the $s$-th positive root of $J_\ell$.
For a dimensionless radial coordinate $x=r/R$, the corresponding
normalized basis is
$\mathcal B_{\ell s}(x):=R B_{\ell s}(Rx)$, so that
$\int_0^1 x\,\dd x\,\mathcal B_{\ell s}(x)
\mathcal B_{\ell t}(x)=\delta_{st}$. This convention is used
for the radial integrals over $x$ below. Defining
\begin{equation}\label{eq:scales}
\begin{aligned}
E_R&=\frac{\hbar^2}{2MR^2},\quad
 a=\frac RL,\quad b=\frac{\Om R^2}{L},\\
X_{st}^{(\ell)}&=\int_0^R r\,\dd r\,B_{\ell s}(r)
\frac{r^2}{R^2}B_{\ell t}(r).
\end{aligned}
\end{equation}
the exact Galerkin matrix is
\begin{equation}\label{eq:matrix}
\frac{\langle s,p|H_m|t,n\rangle}{E_R}
=\begin{aligned}[t]
&\delta_{st}\delta_{pn}
 \bigl[j_{\ell s}^2+(a\pi n)^2\bigr]\\
&+b^2(\pi n)^2X_{st}^{(\ell)}\delta_{pn}\\
&+2\ii mb\,\delta_{st}D_{pn}.
\end{aligned}
\end{equation}
The matrix is Hermitian because its only imaginary part is $\ii D$ with $D^\mathsf T=-D$. Truncation to $N_r$ Bessel functions and $N_z$ axial modes gives variational upper bounds, and $\partial_z$ is never replaced by $\pm\ii k_n$.

\subsection{Exact $m=0$ sector and the radial boundary}\label{sec:m0}
When $m=0$, the cross term in Eq.~\eqref{eq:Hm} vanishes and the axial modes separate exactly. Denoting the physical axial standing-wave number by $n_z$, with $k_{n_z}=n_z\pi/L$, the unbounded-disk energies are the Fock--Darwin levels \citep{Fock1928,Darwin1931} shifted by the axial kinetic energy,
\begin{equation}\label{eq:m0exact}
E_{n_r,0,n_z}^{R=\infty}=\frac{\hbar^2k_{n_z}^2}{2M}
+\frac{\hbar^2|\Om|k_{n_z}}{M}(2n_r+1),\quad n_r=0,1,\ldots.
\end{equation}
The normalized radial eigenstates are those of the two-dimensional oscillator with $\beta=|\Om|k_{n_z}$. At the parameters $L=100\,\mathrm{nm}$, $\Om=5\times10^6\,\mathrm{m}^{-1}$, and $M=0.023m_e$, the exact spacing between successive radial levels in this sector is
\begin{equation}\label{eq:spacing}
\Delta E_{\rm rad}=\frac{2\hbar^2|\Om|k_1}{M}=1.040818\,\mathrm{meV}.
\end{equation}
In the Dirichlet-axial system with $R=\infty$ this is the $m=0$ radial excitation gap between levels of the same $m$; no $m\leftrightarrow-m$ splitting is involved.

If a finite radial wall $R$ is retained, let
$\beta=|\Om|k_{n_z}$; the regular radial solution for $m=0$ is
\begin{equation}
\begin{aligned}
F(r)&=e^{-\beta r^2/2}\,{}_1F_1\!\left(
\frac12-\frac{E-\hbar^2k_{n_z}^2/(2M)}{2\hbar\omega_{n_z}};\right.\\
&\left.\qquad 1;\,\beta r^2\right),
\qquad\omega_{n_z}=\frac{\hbar\beta}{M}.
\end{aligned}
\end{equation}
where the Dirichlet equation $F(R)=0$ determines the energy. For $\Om=0$ this tends to the Bessel-well spectrum $E_R[j_{0s}^2+(a\pi n)^2]$. The finite-radius $m=0$ sector thus provides an exact test of both limits and of the numerical code.

\section{Weak twist, gauge-adapted representation, and radial confinement}\label{sec:weak}
For a small twist the cavity spectrum can be followed analytically. The second-order shift and its change of sign with the aspect ratio are derived in Sec.~\ref{sec:pert}. Section~\ref{sec:gauge} introduces an exact representation built on a radius-dependent axial phase, which we also use in the numerical work, and Sec.~\ref{sec:crossover} connects the finite cylinder with the unbounded Landau problem.

\subsection{Weak-twist theory, Bessel moments, and sign reversal}\label{sec:pert}
For the finite cylinder, write $H_m=H_m^{(0)}+\Om H_m^{(1)}+\Om^2H_m^{(2)}$ with
\begin{equation}
H_m^{(1)}=-\frac{\hbar m}{M}p_z,\quad p_z=-\ii\hbar\partial_z,\qquad
H_m^{(2)}=\frac{r^2p_z^2}{2M}.
\end{equation}
At zero twist, $|s,m,n\rangle=B_{|m|s}\chi_n e^{\ii m\phi}/\sqrt{2\pi}$ and
$E_{smn}^{(0)}=E_R[j_{|m|s}^2+(a\pi n)^2]$. The diagonal expectation of $H_m^{(1)}$ vanishes. Since $H_m^{(1)}$ does not change the radial index, its virtual axial contribution is
\begin{equation}\label{eq:sump}
\sum_{p\ne n}\frac{|\langle p|p_z|n\rangle|^2}{\eps_n-\eps_p}=-\frac M2,
\qquad \eps_n=\frac{\hbar^2k_n^2}{2M}.
\end{equation}
To prove it, note that the phase $e^{\ii\eta z/\hbar}$ maps $(p_z-\eta)^2/(2M)$ unitarily onto $p_z^2/(2M)$ on the Dirichlet domain, so the second-order contribution of $-\eta p_z/M$ must cancel the constant $\eta^2/(2M)$.

The Bessel radial moment also has a closed form:
\begin{equation}\label{eq:C}
\frac{\langle s,m|r^2|s,m\rangle}{R^2}
=C_{\ell s}=\frac13\left[1+\frac{2(\ell^2-1)}{j_{\ell s}^2}\right],\qquad\ell=|m|.
\end{equation}
Combining Eqs.~\eqref{eq:sump} and \eqref{eq:C} gives, for an isolated level near $\Om=0$,
\begin{equation}\label{eq:weak}
\begin{aligned}
E_{smn}(\Om)&=E_{smn}^{(0)}
+\frac{\hbar^2\Om^2}{2M}\biggl[
\biggl(\frac{n\pi R}{L}\biggr)^2\\
&\quad\times C_{|m|s}-m^2\biggr]
+O(\Om^4).
\end{aligned}
\end{equation}
The correction is even in $m$ and leaves the time-reversal pair degenerate. The finite-radius susceptibility is
\begin{equation}\label{eq:chi}
\chi_{smn}^{(R)}:=\left.\frac{\partial^2E_{smn}}{\partial\Om^2}\right|_{0}
=\frac{\hbar^2}{M}\left[(n\pi R/L)^2C_{|m|s}-m^2\right].
\end{equation}
For $|m|=1$, $C_{1s}=1/3$ for every radial root, so that
\begin{equation}\label{eq:critical}
\left(\frac RL\right)_{\!\mathrm{crit}}=\frac{\sqrt3}{\pi n}\qquad (|m|=1).
\end{equation}
For the lowest axial mode $n=1$, this ratio is $0.551328895\ldots$: the quadratic energy shift is negative for $R/L<\sqrt3/\pi$ and positive above it. More generally, the zero of the quadratic response for $m\ne0$ is
\begin{equation}\label{eq:criticalgeneral}
\left(\frac RL\right)_{\!\mathrm{crit}}=
\frac{|m|}{\pi n\sqrt{C_{|m|s}}}.
\end{equation}
The aspect ratio thus controls the sign of the twist response, which has no counterpart in Ref.~\citep{SilvaOQE2026}. In Euclidean waveguides a twist of a noncircular cross section acts repulsively \citep{Ekholm2008,KrejcirikZuazua2010}; here the sign of the leading shift of the $m\neq0$ levels depends on the aspect ratio. The shift applies to the $\pm m$ doublet as a whole, which stays degenerate.

The same result follows from an exact unitary transformation, which also gives a faster-converging variational basis for the local currents. We construct it in Sec.~\ref{sec:gauge}.

\subsection{Exact gauge-adapted Dirichlet representation}\label{sec:gauge}
The square completion of Eq.~\eqref{eq:positive} in Sec.~\ref{sec:crossover} can be implemented as a unitary map on the operator domain. Define the real, radius-dependent axial phase
\begin{equation}\label{eq:gaugephase}
\begin{aligned}
\alpha_m(r)&=\frac{m\Om}{1+\Om^2r^2},\\
u_m(r,z)&=e^{\ii\alpha_m(r)z}v_m(r,z),\\
\alpha_m'(r)&=-\frac{2m\Om^3r}{(1+\Om^2r^2)^2}.
\end{aligned}
\end{equation}
The phase is single-valued on the finite cylinder and preserves Dirichlet traces at $r=R$ and $z=\pm L/2$. Inserted in the form \eqref{eq:form}, it turns the fixed-$m$ kinetic form into
\begin{equation}\label{eq:gaugeform}
\begin{aligned}
Q_m[v]&=\frac{\hbar^2}{2M}\int_0^R r\,\dd r
\int_{-L/2}^{L/2}\dd z\,\biggl\{\\
&\bigl|\bigl(\partial_r+\ii z\alpha_m'(r)\bigr)v\bigr|^2\\
&+(1+\Om^2r^2)|\partial_zv|^2\\
&+\frac{m^2}{r^2(1+\Om^2r^2)}|v|^2
\biggr\}.
\end{aligned}
\end{equation}
The two forms are unitarily equivalent, but the twist is not removed: the radial derivative acquires the term $z\alpha_m'(r)$, whose coefficient is of order $\Om^3$ at fixed $r$, whereas
\begin{equation}\label{eq:gaugeweak}
\begin{aligned}
\frac{m^2}{r^2(1+\Om^2r^2)}
&=\frac{m^2}{r^2}-m^2\Om^2+O(\Om^4),\\
(1+\Om^2r^2)|\partial_zv|^2
&=|\partial_zv|^2+\Om^2r^2|\partial_zv|^2.
\end{aligned}
\end{equation}
For finite $R$, Eqs.~\eqref{eq:gaugeform}--\eqref{eq:gaugeweak} give the whole $O(\Om^2)$ correction of Eq.~\eqref{eq:weak} at first order and show why no linear splitting appears, without assigning Dirichlet eigenvalues to $p_z$. The expansion is not uniform as $R\to\infty$.

For a numerical implementation let $x=r/R$, $\zeta=(z+L/2)/L$, $a=R/L$, $b=\Om R^2/L$, and define
\begin{equation}\label{eq:qdef}
\begin{aligned}
q_m(x)&=\frac{mb}{a^2+b^2x^2},\\
U(x,\zeta)&=e^{\ii q_m(x)(\zeta-1/2)}V(x,\zeta),\\
q_m'(x)&=-\frac{2mb^3x}{(a^2+b^2x^2)^2}.
\end{aligned}
\end{equation}
In units $E_R=\hbar^2/(2MR^2)$, the transformed quadratic form is
\begin{equation}\label{eq:gaugeformdim}
\begin{aligned}
\frac{Q_m[V]}{E_R}&=\int_0^1 x\,\dd x\int_0^1\dd\zeta\,\biggl\{\\
&\left|\left[\partial_x+\ii q_m'(x)
(\zeta-\tfrac12)\right]V\right|^2\\
&+(a^2+b^2x^2)|\partial_\zeta V|^2\\
&+\frac{m^2a^2}{x^2(a^2+b^2x^2)}|V|^2
\biggr\}.
\end{aligned}
\end{equation}
A Bessel--Dirichlet expansion of $V$ gives a second Hermitian Galerkin scheme that contains the radial gauge term exactly (Appendix~\ref{app:gaugematrix}).

The same representation simplifies the physical current. With the normalization scale $j_0=\hbar/(2\pi M R^3L)$ and $\rho=|V|^2$, one finds
\begin{align}\label{eq:gaugecurrents}
\frac{J_{\hat\phi}}{j_0}&=\frac{ma^2\rho}{x(a^2+b^2x^2)}-bx\,\im(V^*\partial_\zeta V),\\
\frac{J^z}{j_0}&=\frac{a^2+b^2x^2}{a}\,\im(V^*\partial_\zeta V),\nonumber\\
\frac{J^{r}}{j_0}&=\im(V^*\partial_xV)+q_m'(x)(\zeta-\tfrac12)\rho.\nonumber
\end{align}
The $-\Om m\rho$ term of $J^z$ cancels exactly. The cancellation comes from the change of representation, and the physical current is unchanged.

By the Hellmann--Feynman theorem \citep{Feynman1939}, any normalized isolated finite-cylinder level satisfies
\begin{equation}\label{eq:HF}
\begin{aligned}
\frac{\dd E_{\alpha m}}{\dd\Om}
=\frac{\hbar^2}{M}\biggl[
&\Om\int r\,\dd r\,\dd z\,r^2|\partial_zu_m|^2\\
&-m\,\im\int r\,\dd r\,\dd z\,
 u_m^*\partial_zu_m\biggr].
\end{aligned}
\end{equation}
We compared Eq.~\eqref{eq:HF} with centered finite differences in the original Galerkin matrix for $m=0,1,2$ and several nonzero twists. The discrepancies were below $2\times10^{-8}$ in the dimensionless derivative $\dd(E/E_R)/\dd b$.

\subsection{Radial confinement, singular limits, and geometric crossover}\label{sec:crossover}
Positivity of Eq.~\eqref{eq:form} and geometric trapping can be seen without a single-subband approximation. In the sector $m$ the axial part of the form can be completed to a square at every fixed radius:
\begin{equation}\label{eq:positive}
\begin{aligned}
&|\partial_zu|^2+
\left|\frac{\ii m}{r}u-\Om r\partial_zu\right|^2\\
&=(1+\Om^2r^2)
\left|\partial_zu-\frac{\ii m\Om}{1+\Om^2r^2}u\right|^2\\
&\quad+\frac{m^2}{r^2(1+\Om^2r^2)}|u|^2.
\end{aligned}
\end{equation}
At fixed $r$ the phase $\exp[\ii m\Om z/(1+\Om^2r^2)]$ preserves the axial Dirichlet data, and the one-dimensional Poincar\'e inequality gives, as in the Hardy-type bounds for twisted waveguides \citep{Ekholm2008,KrejcirikZuazua2010},
\begin{equation}\label{eq:lower}
\int_{-L/2}^{L/2}\dd z\left|\partial_zu-\frac{\ii m\Om}{1+\Om^2r^2}u\right|^2
\ge \frac{\pi^2}{L^2}\int_{-L/2}^{L/2}\dd z\,|u|^2.
\end{equation}
For $\Om\ne0$ this gives the confining lower bound $(\hbar^2/2M)(\pi/L)^2(1+\Om^2r^2)$ for the axial energy density at radius $r$. Because the phase depends on $r$, it cannot remove the coupling globally without generating radial derivatives. The bound explains the radial confinement without diagonalizing $p_z$.

For $R=\infty$ the excitation scale in Eq.~\eqref{eq:m0exact} is proportional to $|\Om|$, whereas at finite $R$ each level starts from an isolated Bessel eigenvalue at $\Om=0$ with the regular expansion \eqref{eq:weak}. The parameter
\begin{equation}\label{eq:xi}
\xi_n=|\Om|k_nR^2=\pi n|b|
\end{equation}
compares the wall radius with the oscillator length $\ell_n=(|\Om|k_n)^{-1/2}$, since $\xi_n=R^2/\ell_n^2$. Figure~\ref{fig:crossover} shows the two regimes. Derivatives at $\Om=0$ taken after $R\to\infty$ differ from the limit $R\to\infty$ of the finite-$R$ derivatives.

For the lowest $m=0$ level a variational bound quantifies the approach to the unbounded limit. Let $E_{0,0,n}^{(R)}$ denote the radial ground-state energy at radius $R$ and $E_{0,0,n}^{(\infty)}$ the oscillator value \eqref{eq:m0exact}. Domain monotonicity and the Gaussian ground state give
\begin{equation}\label{eq:cutoffbound}
\begin{aligned}
0&\le E_{0,0,n}^{(R)}-E_{0,0,n}^{(\infty)}\\
&\le4E_R\frac{e^{-\xi_n/4}}{1-e^{-\xi_n/4}},\\
\xi_n&=|\Om|k_nR^2>0.
\end{aligned}
\end{equation}
To obtain it, multiply the Gaussian by a radial cutoff equal to one for $r\le R/2$ and decreasing linearly to zero at $r=R$. The ground-state representation of the oscillator form leaves an excess proportional to $|\partial_r\chi_R|^2$, and the Gaussian weight beyond $R/2$ is $e^{-\xi_n/4}$ (Appendix~\ref{app:cutoff}). The bound decays exponentially in $\xi_n$ but is not optimal. For fixed $\Om\ne0$ the finite-wall eigenvalue approaches the oscillator value as $R\to\infty$, while the limits $\Om\to0$ and $R\to\infty$ do not commute in the energy expansion.

\begin{figure}[t]
\centering\includegraphics[width=\linewidth]{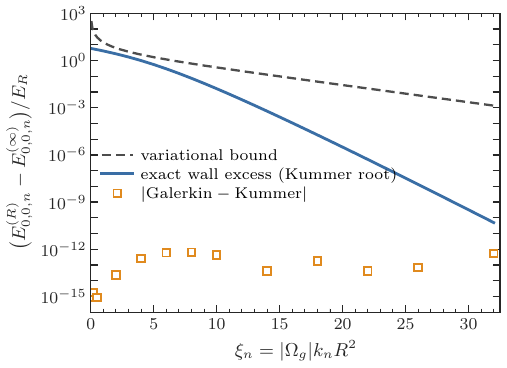}
\caption{Radial-wall correction for $m=0$. Solid line: excess of the finite-$R$ energy over the unbounded oscillator value, from the zero of the confluent hypergeometric solution. Dashed line: bound \eqref{eq:cutoffbound}. Squares: difference between the Galerkin and hypergeometric eigenvalues, at the level of rounding errors (at $\xi_n=1$ both give $\delta=6$ exactly, and the point is omitted).}\label{fig:bound}
\end{figure}

The finite-$R$ eigenvalue for $m=0$ is also given by the first zero of $\,{}_1F_1\big(\tfrac12-\delta/(4\xi_n);1;\xi_n\big)$, where $\delta=(E-E_{z,n})/E_R$. For $0.05\le\xi_n\le32$ this root agrees with the Bessel Galerkin value to better than $3\times10^{-12}$ in units of $E_R$. Figure~\ref{fig:bound} shows the exact wall excess, the bound, and the difference between the two calculations.

\section{Spectra and probability currents}\label{sec:spectra}
We now leave perturbation theory. The spectrum is computed and checked in Sec.~\ref{sec:num}. The probability current in the metric, and the way time reversal pairs the currents of $m$ and $-m$, are derived in Sec.~\ref{sec:currents}, and Sec.~\ref{sec:currentrefinement} tests how well the integrated and local currents converge.

\subsection{Nonperturbative spectra and numerical verification}\label{sec:num}
We diagonalized Eq.~\eqref{eq:matrix} in Bessel$\times$sine bases, with Gauss--Legendre quadrature for $X_{st}$. This is a Hermitian discretization of the original operator that keeps all subband couplings within the basis. As a cross-check for $|m|\ge1$ we also solved the two-dimensional equation for $v_m(r,z)=\sqrt r\,u_m(r,z)$ by finite differences with Dirichlet conditions.

For numerical illustration we take
\begin{equation}\label{eq:params}
\begin{gathered}
R=200\,\mathrm{nm},\quad L=100\,\mathrm{nm},\\
M=0.023m_e,\quad
\Om=5.0\times10^6\,\mathrm{m}^{-1}.
\end{gathered}
\end{equation}
so that $a=b=2$ and $E_R=0.041412849\,\mathrm{meV}$. They set the scale of the effective metric model and say nothing about the atomistic structure of a rolled-up semiconductor. They are the lower ends of the ranges given for rolled-up tubes in Table~1 of Ref.~\citep{SilvaOQE2026} ($R=0.2$--$0.5\,\mu$m, $L=0.1$--$2\,\mu$m). The twist itself calls for a radius larger than the length. For $k=\pi/L$ the harmonic length of the traveling sector is $\ell_{\rm ho}=(\pi\Om/L)^{-1/2}=79.8$ nm (Sec.~\ref{sec:travel}), and the radial confinement produced by the twist dominates the radial wall only when $\xi_1=R^2/\ell_{\rm ho}^2$ of Eq.~\eqref{eq:xi} is large compared with unity (Sec.~\ref{sec:crossover}). With this $\Om$ and $L$ that requires $R\gtrsim L$. Here $\xi_1=2\pi$ and $R\simeq2.5\,\ell_{\rm ho}$, so the cavity is a disk with $R/L=2$, twisted by $\Om L=0.5$ rad across its thickness. A cylinder longer than its radius with the same $\Om$ needs a larger $\Om L$; for $L=1\,\mu$m and $R=500$ nm, for example, $\Om L=5$, $\ell_{\rm ho}=252$ nm, $\xi_1=3.9$ and $E_R$ drops to $6.6\,\mu$eV.

\begin{figure}[t]
\centering\includegraphics[width=\linewidth]{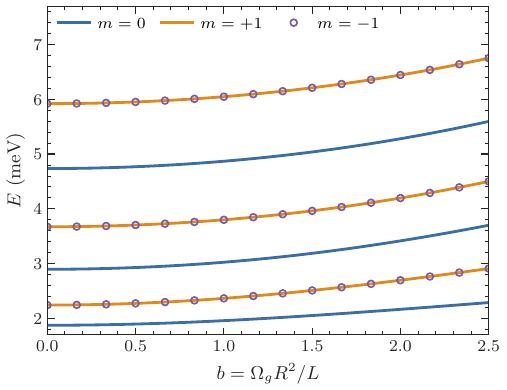}
\caption{Lowest three levels of the finite cylinder in the sectors $m=0$ and $m=+1$ (lines) and $m=-1$ (circles) as functions of $b=\Om R^2/L$, at $a=2$ and with $R$ and $M$ of Eq.~\eqref{eq:params}. The $m=\pm1$ levels coincide for all $b$. Gauge-adapted Galerkin basis with $(N_r,N_z)=(12,16)$.}\label{fig:spectra}
\end{figure}
\begin{table*}[t]
\centering
\caption{Six lowest energies (meV) in each orbital sector for $a=b=2$, from the gauge-adapted Galerkin basis with $N_r=22$, $N_z=26$. The two signs of $m$ agree to eigensolver accuracy. Enlarging the basis to $(30,40)$ changes no displayed digit, and the untransformed basis converges to the same values.}
\label{tab:eigs}
\begin{tabular}{crrrrrr}\toprule
$m$ & 1 & 2 & 3 & 4 & 5 & 6\\\midrule
$0$ &2.165416&3.415503&5.280499&7.580557&7.939137&9.670822\\
$+1$&2.696325&4.197594&6.444278&8.586437&9.505923&10.709769\\
$-1$&2.696325&4.197594&6.444278&8.586437&9.505923&10.709769\\
$+2$&3.226859&5.041325&7.674892&9.533008&11.135985&11.735803\\
$-2$&3.226859&5.041325&7.674892&9.533008&11.135985&11.735803\\\bottomrule
\end{tabular}
\end{table*}
\begin{figure}[t]
\centering\includegraphics[width=\linewidth]{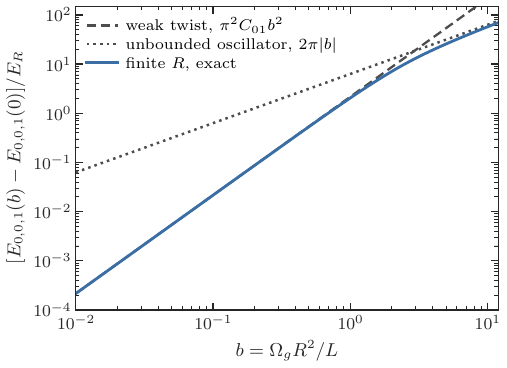}
\caption{Shift of the lowest $m=0$ level, in units of $E_R$, as a function of $b$ at finite $R$ (solid), compared with the weak-twist result $\pi^2C_{01}b^2$ (dashed) and with the energy $2\pi|b|$ of the unbounded radial oscillator (dotted). Because the zero-twist Bessel energy $j_{01}^2E_R$ is subtracted, the solid curve approaches $2\pi|b|-j_{01}^2$ at large $b$.}\label{fig:crossover}
\end{figure}
\begin{figure}[t]
\centering\includegraphics[width=\linewidth]{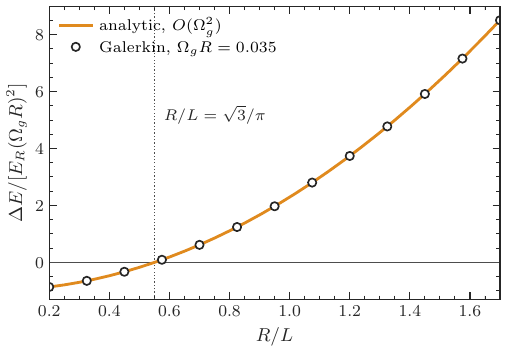}
\caption{Weak-twist coefficient of the $|m|=1$, $s=n=1$ level as a function of $R/L$: analytic $O(\Om^2)$ result (line) and Galerkin energy differences at $\Om R=0.035$ in the gauge-adapted basis (circles). The coefficient changes sign at $R/L=\sqrt3/\pi$; the $m=\pm1$ pair stays degenerate.}\label{fig:suscept}
\end{figure}
For $m=1$ and $b=2$ the original Bessel--sine basis gives $2.69632647$ meV at $(N_r,N_z)=(24,28)$, and the gauge-adapted basis gives $2.69632499$ meV already at $(22,26)$. The two values differ by $1.48\times10^{-6}$ meV and both decrease toward the same variational limit. The finite-difference result is $2.69610343$ meV with $120\times120$ interior points and approaches the Galerkin value from below as the mesh is refined ($2.69626688$ meV with $240\times240$ points); we use it only as a consistency check. Figure~\ref{fig:spectra} shows the time-reversal doublets and Fig.~\ref{fig:suscept} the sign reversal of the weak-twist coefficient. The circles in Fig.~\ref{fig:suscept} were computed in the gauge-adapted basis, which is converged at $N_z=14$. Near the critical ratio the coefficient is small and the original basis converges slowly in $N_z$: at $R/L=\sqrt3/\pi$ and $\Om R=0.035$, the quantity $[E(\Om)-E(0)]/[E_R(\Om R)^2]$ is $7.2\times10^{-4}$ for $N_z=14$ and $3.98\times10^{-4}$ for $N_z=150$, which coincides with the gauge-adapted value and is the higher-order remainder of the expansion.

\subsection{Probability current and time-reversal pairing}\label{sec:currents}
The scalar current follows from the continuity equation in the metric volume measure,
\begin{equation}\label{eq:Jcoord}
J^i=\frac{\hbar}{M}\im\bigl(\Psi^*g^{ij}\partial_j\Psi\bigr),\qquad
\partial_t(r|\Psi|^2)+\partial_i(rJ^i)=0.
\end{equation}
In the orthonormal frame \eqref{eq:frame} its physical components are
\begin{align}\label{eq:Jframe}
J_{\hat r}&=\frac{\hbar}{M}\im(\Psi^*\partial_r\Psi),\\
J_{\hat\phi}&=\frac{\hbar}{M}\im\left[\Psi^*\bigl(r^{-1}\partial_\phi-\Om r\partial_z\bigr)\Psi\right],\\
J_{\hat z}&=\frac{\hbar}{M}\im(\Psi^*\partial_z\Psi),
\end{align}
with $J^\phi=J_{\hat\phi}/r$ and $J^z=J_{\hat z}-\Om rJ_{\hat\phi}$. An eigenstate with $m\ne0$ can carry a local azimuthal circulation although the Hamiltonian is time-reversal invariant, and its partner carries the opposite current:
\begin{equation}\label{eq:JTR}
\mathbf J[\Psi_{-m}]=-\mathbf J[\Psi_m].
\end{equation}
In a mixture that occupies both partners equally, $\mathbf J_{\mathrm{ensemble}}=0$. A current carried by a prepared eigenstate is different from a persistent current in equilibrium, which the scalar Hamiltonian does not produce. The walls enforce $\Psi=0$ and hence $J^z=0$ at $z=\pm L/2$, but axial and azimuthal flow patterns can exist inside.
A stationary eigenstate of definite $m$ also obeys $\partial_\phi\mathbf J=0$. Integrating the continuity equation over a transverse disk, with regularity at the axis and the radial Dirichlet condition, gives
\begin{equation}\label{eq:zero_flux}
\begin{gathered}
\frac{\dd}{\dd z}\int_0^R r\,\dd r\,J^z(r,z)=0,\\
\int_0^Rr\,\dd r\,J^z(r,z)=0
\quad\text{for every }z.
\end{gathered}
\end{equation}
The second equality follows from the vanishing flux at the walls. It concerns the coordinate flux $J^z$ through the surfaces $z=\mathrm{const}$; the local frame component $J_{\hat z}$ is not constrained by it. A local $J^z$ can be compensated by opposite flow elsewhere in the same cross section.

\begin{figure}[t]
\centering\includegraphics[width=\linewidth]{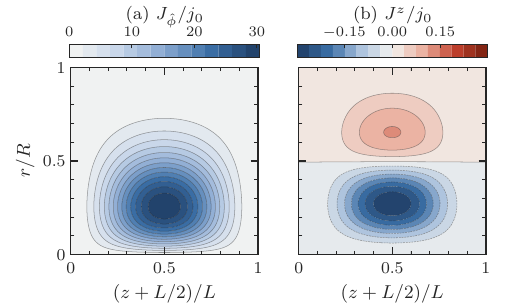}
\caption{Probability current in the lowest $m=+1$ eigenstate at $a=b=2$, computed in the gauge-adapted basis ($N_r=22$, $N_z=26$): (a) orthonormal azimuthal component $J_{\hat\phi}$; (b) coordinate axial component $J^z$. Both are in units of $j_0=\hbar/(2\pi M R^3L)$ for the normalization $\Psi=e^{\ii m\phi}U(r/R,(z+L/2)/L)/(R\sqrt{2\pi L})$, so they are dimensionless probability currents. The $m=-1$ partner has the reversed patterns.}\label{fig:current}
\end{figure}

\subsection{Convergence of the integrated and local currents}\label{sec:currentrefinement}
Energy convergence does not guarantee that approximate eigenstates conserve the current to the same accuracy, because the axial derivative is sensitive to the slowly decaying tail of the sine expansion. Table~\ref{tab:currentaudit} compares the residual of the flux identity \eqref{eq:zero_flux} in the original and the gauge-adapted bases for $m=+1$ at $R/L=b=2$, in units of $j_0$.
\begin{table}[t]
\centering\small
\caption{Lowest $m=1$ energy and residual of the cross-sectional flux, $\max_\zeta|\int_0^1x\,\dd x\,J^z/j_0|$ (maximum over 4001 values of $\zeta$), in the original (std) and gauge-adapted bases. The gauge-adapted truncations satisfy the flux identity to near machine precision.}\label{tab:currentaudit}
\begin{tabular}{rrrrrr}\toprule
$N_r$ & $N_z$ & $E_{\rm std}$ (meV) & $E_{\rm gauge}$ (meV) & Flux$_{\rm std}$ & Flux$_{\rm gauge}$\\\midrule
10&12&2.696341586&2.696325027&$1.47\times10^{-2}$&$6.6\times10^{-11}$\\
14&18&2.696330279&2.696325000&$6.92\times10^{-3}$&$1.4\times10^{-11}$\\
18&22&2.696327967&2.696324994&$4.73\times10^{-3}$&$6.3\times10^{-12}$\\
22&26&2.696326828&2.696324992&$3.44\times10^{-3}$&$3.3\times10^{-12}$\\\bottomrule
\end{tabular}
\end{table}

In the gauge-adapted calculation the largest value of $|J^z|/j_0$ on a fixed interior grid changes from $0.29723$ at $(10,12)$ to $0.29669$ at $(22,26)$. We now examine the pointwise convergence.

Equation~\eqref{eq:zero_flux} constrains only the integrated flux, while a reliable map of the local current needs converged radial and axial derivatives. We therefore compare the gauge-adapted $m=1$ state at $a=b=2$ for several bases $(N_r,N_z)$ with a larger reference basis $(28,32)$, evaluating the three current components on a common interior grid, and define the weighted relative error
\begin{equation}\label{eq:localerror}
\begin{aligned}
\epsilon_J(N_r,N_z)
&=\left[\frac{\int_{\mathcal I}x\,\dd x\,\dd\zeta\,
|J_{N_r,N_z}-J_{28,32}|^2}
{\int_{\mathcal I}x\,\dd x\,\dd\zeta\,|J_{28,32}|^2}
\right]^{1/2},\\
\mathcal I&=[0.035,0.965]\times[0.025,0.975].
\end{aligned}
\end{equation}
Thin strips next to the walls are excluded, because derivatives are hard to resolve on Dirichlet boundaries. Since the reference is itself a finite basis, $\epsilon_J$ measures numerical stability and does not bound the absolute error.

\begin{table*}[t]
\centering\small
\caption{Relative interior errors of the current components, Eq.~\eqref{eq:localerror}, with respect to the gauge-adapted reference $(28,32)$, and the lowest $m=1$ energy. The $z$ entry refers to the coordinate current $J^z$. Uniform $121\times121$ grid on $\mathcal I$.}\label{tab:localcurrent}
\begin{tabular}{rrrrr}\toprule
$(N_r,N_z)$ & $\epsilon_{J^{r}}$ & $\epsilon_{J_{\hat\phi}}$ & $\epsilon_{J^z}$ & $E$ (meV)\\\midrule
$(8,10)$ & $5.966\times10^{-3}$&$7.12\times10^{-5}$&$6.423\times10^{-3}$&$2.696325098$\\
$(12,16)$&$3.308\times10^{-3}$&$1.56\times10^{-5}$&$2.489\times10^{-3}$&$2.696325007$\\
$(20,24)$&$1.527\times10^{-3}$&$3.09\times10^{-6}$&$7.066\times10^{-4}$&$2.696324993$\\
$(24,28)$&$1.212\times10^{-3}$&$1.99\times10^{-6}$&$4.816\times10^{-4}$&$2.696324992$\\\bottomrule
\end{tabular}
\end{table*}
\begin{figure}[t]
\centering\includegraphics[width=\linewidth]{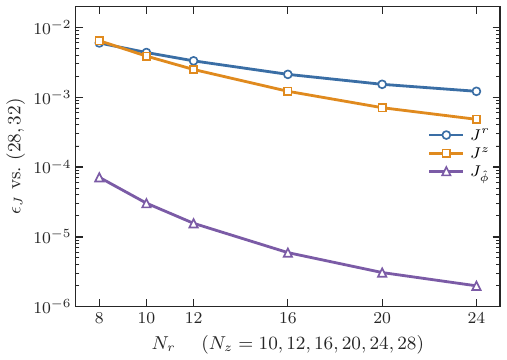}
\caption{Relative interior errors $\epsilon_J$, Eq.~\eqref{eq:localerror}, of the three current components with respect to the $(28,32)$ gauge-adapted reference. The azimuthal component converges much faster than $J^r$ and $J^z$. Thin layers next to the walls are not included in the measure.}\label{fig:localcurrent}
\end{figure}
At $(24,28)$ the interior errors are $0.121\%$ for $J^r$, $0.00020\%$ for $J_{\hat\phi}$ and $0.048\%$ for $J^z$. We have not examined the maximum error near the walls or the current pattern for $R\to\infty$.

\section{Dipole spectroscopy}\label{sec:spectro}
The pairing of Sec.~\ref{sec:dir} rules out a splitting between $m$ and $-m$, so the effects of the twist have to be looked for elsewhere, and optical transitions are the natural place. We derive an exact longitudinal sum rule in Sec.~\ref{sec:dipole}. In Sec.~\ref{sec:anticross} we find an avoided crossing within one $m$ sector whose gap is cubic in the twist, and in Sec.~\ref{sec:spectroscopy} we give the circular-polarization selection rules and the transitions that the twist activates.

\subsection{Axial dipole response and an exact geometric sum rule}\label{sec:dipole}
In the scalar effective-mass model, let $X^i$ denote multiplication by a Cartesian coordinate, $X^i\in\{x,y,z\}$. For Dirichlet states the double-commutator identity of the Thomas--Reiche--Kuhn sum rule \citep{Thomas1925,ReicheThomas1925,Kuhn1925} reads
\begin{equation}\label{eq:TRKcomm}
[X^i,[H,X^j]]=\frac{\hbar^2}{M}g^{ij},
\end{equation}
where $g^{ij}$ is the inverse metric expressed in those coordinates. In particular, the Cartesian form of Eq.~\eqref{eq:metric}, $\dd s^2=\dd x^2+\dd y^2+[\dd z+\Om(x\dd y-y\dd x)]^2$, gives $g^{xx}=g^{yy}=1$ and $g^{zz}=1+\Om^2r^2$. Completeness of the bound-state spectrum implies the longitudinal dipole sum rule
\begin{equation}\label{eq:TRKz}
S_z(i):=\frac{2M}{\hbar^2}\sum_f(E_f-E_i)|\langle f|z|i\rangle|^2
=1+\Om^2\langle i|r^2|i\rangle.
\end{equation}
Likewise $S_x(i)=S_y(i)=1$. Equation~\eqref{eq:TRKz} is an identity of the scalar metric Hamiltonian. It can be tested only in a system that realizes both this kinetic operator and the coordinate-dipole coupling.

The axial coordinate $z$ is independent of $\phi$, so axial polarization obeys $\Delta m=0$ exactly. At zero twist its matrix elements connect opposite Dirichlet parities:
\begin{equation}\label{eq:zselection}
\langle\chi_p|z|\chi_n\rangle=
\begin{cases}
-\dfrac{8Lpn}{\pi^2(p^2-n^2)^2},&p\neq n,\quad p+n\text{ odd},\\[5pt]
0,&p+n\text{ even},
\end{cases}
\end{equation}
with $z$ measured from the box center. The twist-induced mixing in Eq.~\eqref{eq:matrix} redistributes these oscillator strengths among the eigenstates. The circular dipoles $x\pm\ii y=r e^{\pm\ii\phi}$ obey $\Delta m=\pm1$ and, at zero twist, conserve the axial index. Testing them does not require any orbital splitting.

\begin{figure}[t]
\centering\includegraphics[width=\linewidth]{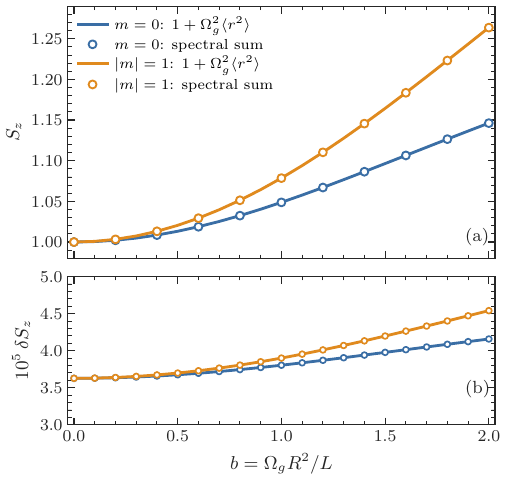}
\caption{(a) Longitudinal dipole sum rule for the lowest state in the $m=0$ and $|m|=1$ sectors at $a=2$: double-commutator value $1+\Om^2\langle r^2\rangle$ (lines) and spectral sum over all states of the $(N_r,N_z)=(14,30)$ basis (circles). (b) Deficit $\delta S_z=1+\Om^2\langle r^2\rangle-S_z$, below $5\times10^{-5}$, caused by the truncated high-energy tail.}\label{fig:sumrule}
\end{figure}

\subsection{Same-orbital anticrossings and a cubic geometric gap}\label{sec:anticross}
The Dirichlet Hamiltonian is block diagonal in $m$ but not in the axial index, so avoided crossings between levels of the same symmetry \citep{vonNeumannWigner1929} can occur within a fixed $m$ sector while $E_m=E_{-m}$ still holds. A nonzero $D_{pn}$ does not by itself produce a first-order gap between states with different radial indices: the order-$b$ term of the original basis is diagonal in the radial index, so its matrix element vanishes when the radial indices differ, even for opposite axial parities. A degenerate expansion to higher order is needed.

Consider the positive-twist finite cylinder with dimensionless parameters $a=R/L$ and $b=\Om R^2/L$ introduced in Eq.~\eqref{eq:scales}. In the unperturbed $|m|=1$ spectrum, two states $A=(s,n)=(1,2)$ and $B=(2,1)$ cross at
\begin{equation}\label{eq:across}
\begin{aligned}
a_0&=\left[\frac{j_{1,2}^2-j_{1,1}^2}{3\pi^2}\right]^{1/2}
=1.080011976\ldots,\\
E_A^{(0)}&=E_B^{(0)}.
\end{aligned}
\end{equation}
The gauge-adapted form \eqref{eq:gaugeform} shows the order of their coupling directly. With $x=r/R$, $\zeta=(z+L/2)/L$, and $q_m(x)=mb/(a^2+b^2x^2)$, its radial connection has the expansion
\begin{equation}\label{eq:qprimeexpand}
 q_m'(x)=-\frac{2mb^3}{a^4}x+O(b^5).
\end{equation}
Let $\mathcal B_s(x)=\mathcal B_{1s}(x)$ denote the
unit-normalized dimensionless radial Bessel functions, and let $s_n(\zeta)=\sqrt2\sin(n\pi\zeta)$ and $Z_{pn}=\int_0^1s_p(\zeta)(\zeta-1/2)s_n(\zeta)\,\dd\zeta$. For the crossing pair, the first nonvanishing direct off-diagonal entry in the transformed basis is
\begin{equation}\label{eq:cubicoffdiag}
\begin{aligned}
\frac{\langle A|H_m^{\rm(gauge)}|B\rangle}{E_R}
&=\ii b^3 C_{AB}+O(b^5),\\
C_{AB}&=-\frac{2m}{a_0^4}Z_{21}
\int_0^1x^2(\mathcal B_1'\mathcal B_2-\mathcal B_1\mathcal B_2')\,\dd x,\\
Z_{21}&=-\frac{16}{9\pi^2}.
\end{aligned}
\end{equation}
Indirect contributions involving the $O(b^2)$ radial term enter the off-diagonal element only beyond third order, because that term preserves $n$. The diagonal shifts move the crossing to
\begin{equation}\label{eq:acrossshift}
 a_\ast^2(b)=a_0^2-\frac{b^2}{3}+O(b^4),
\end{equation}
where we used $\langle x^2\rangle_{|m|=1,s}=1/3$ for every radial mode. Minimizing the separation of the pair over $a$ at fixed $b$ gives the avoided crossing \citep{vonNeumannWigner1929} with the cubic gap
\begin{equation}\label{eq:cubicgap}
\begin{aligned}
\frac{\Delta_{\rm gap}(b)}{E_R}
&=2|C_{AB}|\,|b|^3+O(|b|^5)\\
&=1.6487752187\ldots\,|b|^3+O(|b|^5).
\end{aligned}
\end{equation}
The coefficient follows from a one-dimensional Bessel integral, without diagonalization, and the sectors $m=\pm1$ have identical gaps. For $m=0$ the axial modes separate exactly, states with different $n$ stay uncoupled, and true crossings can remain when the aspect ratio is tuned. The cubic law refers to this isolated radial--axial crossing and to the weak-twist scan described above; other resonances need their own analysis.

\begin{table}[t]
\centering\small
\caption{Lowest avoided gap in the $m=1$ sector, minimized over $a$ at fixed $b$ (gauge-adapted basis, $(N_r,N_z)=(24,32)$). As $b\to0$ the minimum moves to $a_0$ of Eq.~\eqref{eq:across}. Energies use $E_R$ of Eq.~\eqref{eq:params}; at fixed $R$, keeping $b$ fixed while varying $a$ requires $\Om=b/(aR)$.}
\label{tab:gaps}
\begin{tabular}{rrrr}\toprule
$b$&$a_\ast$&$\Delta_{\rm gap}$ (meV)&$\Delta_{\rm gap}/[2E_R|C_{AB}|b^3]$\\\midrule
$0.12$&$1.077792$&$0.00011794$&$0.99959$\\
$0.30$&$1.066229$&$0.00183816$&$0.99707$\\
$0.60$&$1.026254$&$0.01450435$&$0.98344$\\
$0.80$&$0.987246$&$0.03363582$&$0.96213$\\
$1.30$&$0.866195$&$0.12559338$&$0.83722$\\\bottomrule
\end{tabular}
\end{table}
\begin{figure}[t]
\centering\includegraphics[width=\linewidth]{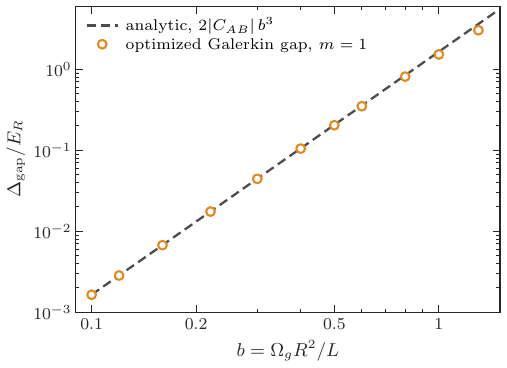}
\caption{Minimized $m=1$ gap from the Galerkin matrix (circles) and the cubic law \eqref{eq:cubicgap} (dashed). The relative difference is $4.1\times10^{-4}$ at $b=0.12$ and grows with $b$ as higher orders contribute. The gap is given in units of the fixed scale $E_R$.}\label{fig:cubic}
\end{figure}

\begin{figure}[t]
\centering\includegraphics[width=\linewidth]{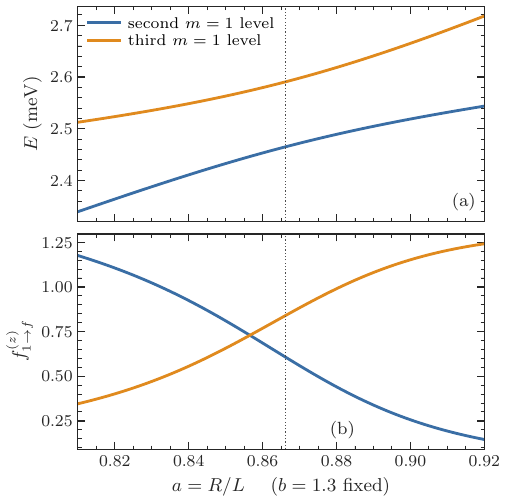}
\caption{Avoided crossing in the $m=1$ sector at fixed $b=1.3$ as the aspect ratio $a=R/L$ is varied (with $\Om=b/(aR)$). (a) Second and third levels. (b) Longitudinal oscillator strengths $f^{(z)}=(E_f-E_i)|\langle f|z|i\rangle|^2/(E_RR^2)$ from the lowest $m=1$ state to these levels, Eq.~\eqref{eq:fpolar}. The dotted line marks the gap minimum, $a_\ast=0.866195$.}\label{fig:anticross}
\end{figure}

Across an isolated resonance the two eigenvectors exchange their radial--axial character, and the axial dipole strength carried by one branch passes to the other. Figure~\ref{fig:anticross} shows the energies and the oscillator strengths from the lowest $m=1$ state to the second and third $m=1$ levels, computed from the full gauge-adapted eigenvectors, without a two-level reduction. In this example $b=1.3$ and the minimum gap of the ideal scalar cavity is $0.12559$ meV.

\subsection{Circular-polarization selection rules and activated transitions}\label{sec:spectroscopy}
Rotational invariance gives azimuthal selection rules that hold at any twist:
\begin{equation}\label{eq:polselection}
 z:\quad\Delta m=0;\qquad d_\pm=x\pm\ii y=r e^{\pm\ii\phi}:\quad\Delta m=\pm1.
\end{equation}
At $b=0$ the transverse operators $d_\pm$ also conserve the axial index $n$, and the centered dipole $z$ connects opposite axial parities. These additional rules fail at finite $b$, where the eigenstates mix axial and radial modes. We characterize the transitions by the dimensionless strengths
\begin{equation}\label{eq:fpolar}
\begin{aligned}
f_{if}^{(z)}
&=\frac{2M(E_f-E_i)}{\hbar^2}
|\langle f|z|i\rangle|^2,\\
f_{if}^{(\pm)}
&=\frac{2M(E_f-E_i)}{\hbar^2}
|\langle f|d_\pm|i\rangle|^2.
\end{aligned}
\end{equation}
Here $d_\pm$ are unnormalized circular coordinates; with unit circular polarization vectors, which carry a factor $1/\sqrt2$, the strengths are halved. They are nonnegative for transitions to higher levels.

Take the nondegenerate $m=0$ ground state as initial state. At zero twist $d_+$ excites the $m=1$ state $(s,n)=(1,1)$ but not the state $(s,n)=(1,2)$. In the gauge-adapted representation the matrix element contains the relative phase
\begin{equation}\label{eq:relativephase}
\begin{aligned}
&\langle f,m+1|d_+|i,m\rangle\\
&\sim\int_0^1x^2\dd x\int_0^1\dd\zeta\,
 v_f^*(x,\zeta)v_i(x,\zeta)\\
&\quad\times\exp\!\left[
-\frac{\ii b(\zeta-1/2)}{a^2+b^2x^2}\right].
\end{aligned}
\end{equation}
where $\sim$ omits only normalization factors. At weak twist the linear term of the phase couples opposite axial parities through $Z_{pn}$, which gives an amplitude proportional to $b$ and
\begin{equation}\label{eq:activated}
\begin{gathered}
f_{0\to(1,2)}^{(+)}(b)=\mathcal A(a)\,b^2+O(b^4),\\
\mathcal A(2)=0.0559408010\ldots.
\end{gathered}
\end{equation}
The coefficient $\mathcal A(a)$ follows from ordinary perturbation theory in the Bessel--sine basis as a single radial overlap (Appendix~\ref{app:activationaudit}); for $a=2$ it equals $0.0559408010\ldots$ in the convention of Eq.~\eqref{eq:fpolar}. The activated intensity does not distinguish $m=+1$ from $m=-1$: the line excited by $d_-$ is degenerate with the $d_+$ line and has the same strength for the $m=0$ initial state.

For the parameters \eqref{eq:params} ($a=b=2$) the dimensionless circular strengths are $f_+^{\rm bright}=1.86328$ and $f_+^{\rm activated}=0.10637$, at excitation energies $0.53091$ meV and $6.42102$ meV. Their ratio, about $5.71\%$, vanishes at $b=0$; at $b=0.15$ and $0.30$ the activated strength is $0.001252$ and $0.004932$, in line with the quadratic onset.

\begin{figure}[t]
\centering\includegraphics[width=\linewidth]{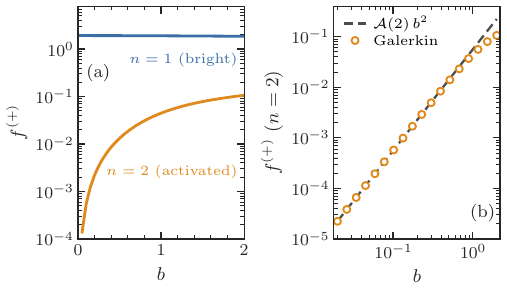}
\caption{(a) Circular-dipole strengths $f^{(+)}$ at $a=2$ from the $m=0$ ground state to the $m=1$ levels connected at zero twist to $(s,n)=(1,1)$ (bright) and $(1,2)$ (activated). (b) At small $b$ the activated strength (circles) follows $\mathcal A(2)b^2$ (dashed). Bessel--sine basis with $(N_r,N_z)=(14,20)$.}\label{fig:activated}
\end{figure}

A second exact sum rule applies to the circular dipoles. For an $m=0$ initial state, rotational invariance and time reversal make the sums into $m_f=+1$ and $m_f=-1$ equal, and combining $[x,[H,x]]=\hbar^2/M$, $[y,[H,y]]=\hbar^2/M$ and $[x,[H,y]]=0$ gives
\begin{equation}\label{eq:sumplus}
\begin{aligned}
S_+(i)&=\frac{2M}{\hbar^2}\sum_{f:\,m_f=+1}
(E_f-E_i)\\
&\quad\times|\langle f|x+\ii y|i,m_i=0\rangle|^2=2,\\
S_-(i)&=2.
\end{aligned}
\end{equation}
The factor two comes from the $x+\ii y$ convention. We checked Eq.~\eqref{eq:sumplus} in the untransformed Bessel--sine basis with its full truncated eigensystem. For $(N_r,N_z)=(16,24)$ the circular sum is $1.999932$ at $b=0$ and $1.999986$ at $b=2$; the deficit comes from the truncated radial basis. The two bases also agree on the activated line, with $0.10637098$ (untransformed) and $0.10637081$ (gauge-adapted) at $b=2$ and $(14,22)$, so the effect depends neither on the phase convention nor on the truncation.

A broadened model spectrum can be written as
\begin{equation}\label{eq:lineprofile}
 \mathscr S_+(\varepsilon)=\sum_f f_{if}^{(+)}
 \frac{\gamma^2}{(\varepsilon-E_f+E_i)^2+\gamma^2}.
\end{equation}
With $\gamma=0.055$ meV, for example, the activated line is resolved. The width $\gamma$ is put in by hand and does not come from a phonon, disorder or radiative calculation; Eq.~\eqref{eq:lineprofile} does not predict absolute absorption.

\section{Physical realization and generalizations}\label{sec:realization}
Here we ask what a real structure would need in order to follow the model, and what changes when its assumptions are relaxed. Section~\ref{sec:implementation} gives the conditions for a physical realization and the measurements that could detect the effects of Sec.~\ref{sec:spectro}. Section~\ref{sec:general} treats other axial boundary conditions, a twist that varies along the axis, and an anisotropic cross section.

\subsection{Conditions for a physical realization and experimental tests}\label{sec:implementation}
The scalar kinetic model does not describe a mechanically twisted cylinder or a thin rolled-up shell. Electrons in a thin semiconductor layer are governed by surface curvature, strain, spin--orbit terms and band offsets, and need not experience the three-dimensional inverse metric used here. A realization must instead reproduce the effective kinetic operator $H$ of Eq.~\eqref{eq:H} itself.

For constant $\Om$, write the metric in Cartesian coordinates \citep{Scott1983,GordonWilson1986} as $\dd s^2=\dd x^2+\dd y^2+[\dd z+\Om(x\dd y-y\dd x)]^2$. Its inverse matrix and determinant are
\begin{equation}\label{eq:tensor}
\begin{aligned}
g^{ij}(x,y)&=\begin{pmatrix}
1&0&\Om y\\
0&1&-\Om x\\
\Om y&-\Om x&1+\Om^2(x^2+y^2)
\end{pmatrix},\\
\det g^{ij}&=1.
\end{aligned}
\end{equation}
Consequently, a scalar effective kinetic energy of the divergence form \citep{BenDanielDuke1966,MorrowBrownstein1984,Burt1992,Foreman1995}
\begin{equation}\label{eq:physicalmapping}
\begin{aligned}
H_{\rm eff}&=-\frac{\hbar^2}{2}\sum_{i,j}\partial_i
\bigl[(M^{-1})^{ij}(\mathbf x)\partial_j\bigr],\\
(M^{-1})^{ij}&=M^{-1}g^{ij}(\mathbf x).
\end{aligned}
\end{equation}
with matching boundary conditions reproduces the Laplace--Beltrami model in Cartesian coordinates, because $\sqrt g=1$. The inverse-mass matrix is symmetric and positive definite. Reproducing it, including the mixed $xz$ and $yz$ entries and the operator ordering \citep{BenDanielDuke1966,MorrowBrownstein1984}, is a much stronger requirement than fabricating a helicoidal shape. For an electronic nanoroll, $k\!\cdot\!p$ or atomistic calculations would have to establish this tensor \citep{Burt1992} and estimate the additional curvature, strain, spin--orbit and interface terms; we have not attempted such a derivation.

That the requirement is not trivial can be seen without coordinates. The one-form
\begin{equation}\label{eq:nonintegrability}
 \beta=\dd z+\Om(x\dd y-y\dd x),\qquad
 \dd\beta=2\Om\,\dd x\wedge\dd y
\end{equation}
is not closed, and hence not exact, for nonzero uniform twist. Together with $\mathcal R=-2\Om^2\neq0$, this means that no change of coordinates applied to a homogeneous isotropic mass tensor can produce the scalar metric. In a local orthonormal cylindrical basis the principal values of the inverse mass, in units of $M^{-1}$, are
\begin{equation}\label{eq:principalvalues}
\begin{gathered}
 \lambda_0=1,\\
 \lambda_{\pm}=1+\frac{t^2}{2}
 \pm\frac{t}{2}\sqrt{t^2+4},\qquad t=|\Om|r,\\
 \lambda_+\lambda_-=1.
\end{gathered}
\end{equation}
For $|\Om|R=1$ the ratio of the extreme principal values at the wall is $\lambda_+/\lambda_-\simeq6.8541$, and the principal axes rotate around the cylinder. The semiconductor parameters of Eq.~\eqref{eq:params} only illustrate the effective-mass scale; they are not derived from the band structure of a curved nanoroll.
\begin{figure}[t]
\centering\includegraphics[width=\linewidth]{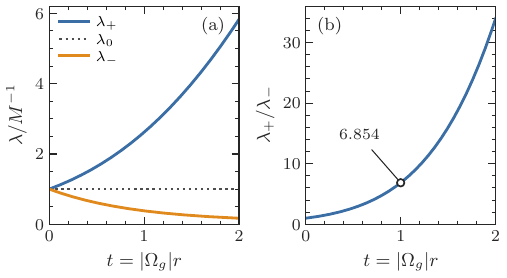}
\caption{(a) Principal values of the inverse-mass tensor, in units of $M^{-1}$, required to reproduce the helicoidal kinetic operator, and (b) their ratio $\lambda_+/\lambda_-$, as functions of $t=|\Om|r$. These are the values an analog simulator would have to implement; they are not material parameters.}\label{fig:tensorcontrast}
\end{figure}

A finite barrier also differs from the Dirichlet model. For an envelope Hamiltonian $H=-\frac{\hbar^2}{2}\partial_i[A^{ij}(\mathbf x)\partial_j]+V$, current conservation at an ideal interface of this divergence-form model requires continuity of the envelope and of $n_iA^{ij}\partial_j\psi$. For the helicoidal tensor the axial conormal derivative is $(1+\Om^2r^2)\partial_z-\Om\partial_\phi$. This rule holds for the single-band operator considered here; atomically abrupt multiband heterojunctions need a separate treatment \citep{Foreman1995,Pistol1999}. Spin--orbit terms can also mix orbital sectors, and in a time-reversal-invariant spinful model the protected partners are Kramers pairs, which need not coincide with the fixed-spin $\pm m$ states used here. The size of these terms matters for the observables below. With the Rashba parameter $\alpha_R\approx3\times10^{-12}$ eV\,m adopted in Ref.~\citep{SilvaOQE2026}, and keeping in mind that $\alpha_R$ in InAs wells depends on gate voltage and well design \citep{Grundler2000}, the splitting $2\alpha_Rk$ is $0.075$ meV for $k=1/\ell_{\rm ho}$ and $0.19$ meV for $k=\pi/L$. This exceeds $E_R=0.041$ meV and is comparable to or larger than every anticrossing gap of Table~\ref{tab:gaps}, so a spinful treatment would be needed before the anticrossings of Sec.~\ref{sec:anticross} could be identified in an InAs structure.

The predictions that an experiment could test are the $m\leftrightarrow-m$ pairing, the zero of the susceptibility as a function of aspect ratio, the cubic anticrossing obtained by retuning the aspect ratio, and the redistribution of axial and transverse dipole strengths. Intraband transitions of laterally confined electrons in the far infrared have been resolved by magnetospectroscopy \citep{SikorskiMerkt1989}. Table~\ref{tab:freq} lists the corresponding energies for the parameters \eqref{eq:params}. Absolute absorption, linewidths, disorder, contact leakage and fabrication tolerances would require a microscopic model. A programmable lattice or an electrical network could approximate a discretized version of the operator, but a mapping onto a rectangular grid does not automatically keep the angular selection rules of the cylinder; we leave this to future work.

Reference~\citep{SilvaOQE2026} proposed three experimental systems with the parameter ranges of its Table~1. Table~\ref{tab:platforms} restates them in terms of the quantities that control the corrected model. For electronic nanorolls the energy scales reach the sub-meV range, but a realization requires the inverse-mass tensor of Eq.~\eqref{eq:tensor} in addition to a helicoidal shape. For photonic waveguides the paraxial equation \eqref{eq:paraxialcorrect} sets a harmonic length $\ell=(\kappa|\Om|)^{-1/2}$ of a few micrometers and rotation angles $|\Om|L$ of $50$--$2000$ rad. For ultracold atoms we know of no scheme that imposes the metric on atomic motion; if one were found, the cavity scale $E_R/h$ for $R=30$--$50\,\mu\mathrm{m}$ would be below $1$ Hz for both $^{87}$Rb and $^{6}$Li.

The detection schemes proposed in Ref.~\citep{SilvaOQE2026} must also be reinterpreted. Spectroscopy of discrete levels, by tunneling, far-infrared absorption or photoluminescence, cannot show a $\pm m$ doublet in the closed cavity, because Theorem~\ref{thm:TR} forbids it; it can show the lines of Table~\ref{tab:freq} and the anticrossings. Magnetometry of persistent currents gives no signal in equilibrium, since the ensemble current vanishes; a signal requires a prepared state of definite $m$, with the sign fixed by Eq.~\eqref{eq:JTR}. In the traveling sector the circulation has the sign of $-\Om k$, Eq.~\eqref{eq:Lkin}, so it reverses with the axial flow. In photonic structures the orbital-angular-momentum spectrum of a centered input is conserved by Eq.~\eqref{eq:paraxialcorrect}; the measurable signatures are the breathing of the beam and the rotation of non-axisymmetric patterns discussed in Sec.~\ref{sec:opt}, the latter subject to the frame dependence discussed there.
\begin{table*}[t]
\centering\small
\caption{Transition and anticrossing energies and frequencies $E/h$ for the effective-mass scale of Eq.~\eqref{eq:params}. The anticrossing values come from the scan in $(a,b)$ of Table~\ref{tab:gaps} and refer to cavities different from the one of the two circular transitions.}\label{tab:freq}
\begin{tabular}{lrr}\toprule
Transition & Energy (meV)&Frequency (THz)\\\midrule
Bright circular $m=0\to+1$, $a=b=2$&$0.53091$&$0.12837$\\
Activated circular $m=0\to+1$, $a=b=2$&$6.42102$&$1.55260$\\
Anticrossing $b=0.8$ at optimized $a$&$0.03364$&$0.00813$\\
Anticrossing $b=1.3$ at optimized $a$&$0.12559$&$0.03037$\\\bottomrule
\end{tabular}
\end{table*}
\begin{table*}[t]
\centering\small
\caption{Experimental systems and parameter ranges of Table~1 of Ref.~\citep{SilvaOQE2026}, restated for the corrected model. $E_R=\hbar^2/(2MR^2)$, with $M=0.023m_e$ for the nanorolls, is the cavity scale and $\hbar\omega_c$ the cyclotron energy of the traveling state with $k=\pi/L$; $\ell=(\kappa|\Om|)^{-1/2}$ is the harmonic length of Eq.~\eqref{eq:paraxialcorrect} for $\lambda_0=1.55\,\mu$m and $n_0=1.45$. For the photonic case $L$ is a propagation length and no closed cavity is involved.}\label{tab:platforms}
\begin{tabular}{llllllp{0.41\textwidth}}\toprule
System & $R$ & $L$ & $\Om$ (m$^{-1}$) & $b=\Om R^2/L$ & $|\Om|R$ & Scale in the corrected model\\\midrule
Nanorolls & $0.2$--$0.5\,\mu$m & $0.1$--$2\,\mu$m & $10^6$--$10^7$ & $0.02$--$25$ & $0.2$--$5$ & $E_R=6.6$--$41\,\mu$eV; $\hbar\omega_c=0.010$--$2.1$ meV\\
Photonics & $20$--$50\,\mu$m & $5$--$20$ mm & $10^4$--$10^5$ & --- & $0.2$--$5$ & $\ell=1.3$--$4.1\,\mu$m; rotation $|\Om|L=50$--$2000$ rad\\
Cold atoms & $30$--$50\,\mu$m & $0.5$--$1$ mm & $10^5$--$10^6$ & $0.09$--$5$ & $3$--$50$ & no known realization; $E_R/h=0.02$--$0.07$ Hz ($^{87}$Rb), $0.3$--$0.9$ Hz ($^{6}$Li)\\\bottomrule
\end{tabular}
\end{table*}

\subsection{Axial boundary topology, phases, and nonuniform twist}\label{sec:general}
The results above assume Dirichlet walls, a constant twist and a circular cross section. The three cases below relax one assumption at a time.

\subsubsection{Periodic and phase-twisted conditions}
For a periodic axis, $\chi_j(z)=L^{-1/2}e^{\ii k_jz}$ with $k_j=2\pi j/L$ and $j\in\mathbb Z$, the traveling solutions of Sec.~\ref{sec:travel} apply. At fixed nonzero $k_j$ the energies of $+m$ and $-m$ differ, but the time-reversed partner $(-m,-k_j)$ is also in the spectrum. A twisted condition $\Psi(z+L)=e^{\ii\theta}\Psi(z)$ shifts $k_j=(2\pi j+\theta)/L$, and time reversal maps it to the problem with phase $-\theta$. Unless an applied flux is reversed at the same time, only $\theta=0$ and $\theta=\pi$ are invariant modulo $2\pi$. A generic phase is an additional physical ingredient and does not follow from the metric. This removes the apparent conflict between the $m$ asymmetry of traveling waves and the pairing in the hard-wall cavity.

\subsubsection{Axially dependent metric twist}
The model extends to a real smooth function $\Om(z)$ in the one-form of Eq.~\eqref{eq:metric}. The determinant is still $g=r^2$, and the Laplace--Beltrami operator becomes
\begin{equation}\label{eq:Omz}
\begin{aligned}
\Delta_{\Om(z)}={}&\frac1r\partial_r(r\partial_r)
+\frac1{r^2}\partial_\phi^2\\
&+(1+\Om(z)^2r^2)\partial_z^2\\
&+2\Om(z)\Om'(z)r^2\partial_z\\
&-2\Om(z)\partial_\phi\partial_z
-\Om'(z)\partial_\phi.
\end{aligned}
\end{equation}
The terms with $\partial_\phi$ contain the symmetrically ordered axial derivative $2\Om\partial_z+\Om'$, whose diagonal element in any real hard-wall mode $\chi_n$ vanishes:
\begin{equation}\label{eq:Omz0}
\int_{-L/2}^{L/2}\dd z\,\chi_n(2\Om\partial_z+\Om')\chi_n
=\bigl[\Om(z)\chi_n^2(z)\bigr]_{-L/2}^{L/2}=0.
\end{equation}
The operator is still real, so Theorem~\ref{thm:TR} holds for a rotationally symmetric boundary. The separation of the $m=0$ sector (Sec.~\ref{sec:m0}) is lost, because the axial coefficient now depends on both $r$ and $z$. For the same reason the uniform-twist formulas cannot be extended by inserting an axially averaged twist $\bar\Om_n$, as was done in Ref.~\citep{SilvaOQE2026}.

\begin{figure}[t]
\centering\includegraphics[width=\linewidth]{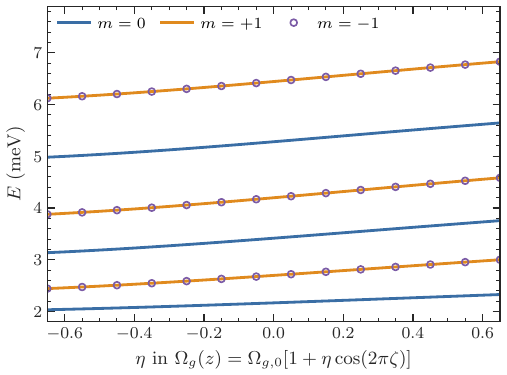}
\caption{Lowest three levels for the nonuniform twist \eqref{eq:OmegaProfile} at $a=b_0=2$, in the sectors $m=0$ and $m=+1$ (lines) and $m=-1$ (circles), from Eq.~\eqref{eq:nonuniform_h} with $(N_r,N_z)=(12,18)$. The matrix contains $b(\zeta)^2$ and the symmetrically ordered derivative $2b\partial_\zeta+b'$.}\label{fig:nonuniform}
\end{figure}

For the numerical treatment we set $\zeta=(z+L/2)/L$, $b(\zeta)=R^2\Om(z)/L$, and $s_n(\zeta)=\sqrt2\sin(n\pi\zeta)$. The dimensionless axial matrices are
\begin{align}\label{eq:nonuniform_matrix}
V_{pn}&=\int_0^1\!\dd\zeta\,b(\zeta)^2 s_p'(\zeta)s_n'(\zeta),\\
C_{pn}&=\int_0^1\!\dd\zeta\,s_p(\zeta)\bigl[2b(\zeta)s_n'(\zeta)+b'(\zeta)s_n(\zeta)\bigr],\nonumber
\end{align}
where $V_{pn}=V_{np}$ and $C_{pn}=-C_{np}$. The complete Galerkin operator is
\begin{equation}\label{eq:nonuniform_h}
\begin{aligned}
\frac{H_m[\Om(z)]}{E_R}&=\mathrm{diag}(j_{|m|s}^2)\otimes I_z\\
&+I_r\otimes\mathrm{diag}[(a\pi n)^2]\\
&+X^{(|m|)}\otimes V+\ii m I_r\otimes C.
\end{aligned}
\end{equation}
Equation~\eqref{eq:nonuniform_h} keeps all $\Om'(z)$ and $\Om(z)^2$ contributions within the basis and reduces to Eq.~\eqref{eq:matrix} for constant twist. As an example we take
\begin{equation}\label{eq:OmegaProfile}
\begin{aligned}
\Om(z)&=\Omega_{g,0}\left[1+\eta\cos\!\left(
\frac{2\pi(z+L/2)}{L}\right)\right],\\
|\eta|&\le0.65.
\end{aligned}
\end{equation}
At $a=b_0=2$, with $b_0=\Omega_{g,0}R^2/L$, the lowest $m=0$ energy changes from $2.03225$ meV at $\eta=-0.65$ to $2.32663$ meV at $\eta=+0.65$, and the $|m|=1$ pair moves from $2.44127$ to $2.99904$ meV while remaining degenerate. These values were obtained with $(N_r,N_z)=(12,18)$; enlarging the basis to $(24,48)$ changes them by less than $1\times10^{-5}$ meV. Figure~\ref{fig:nonuniform} shows the levels as functions of $\eta$.
\subsubsection{Anisotropic cross section}
Reference~\citep{SilvaOQE2026} also considered a transverse anisotropy, $r^2\dd\phi^2\to r^2[1+\delta a(\phi,z)]\dd\phi^2$. Such a term keeps the operator real, so the spectrum remains time-reversal invariant, but it breaks rotational invariance and $m$ is no longer conserved. A component $\delta a\propto\cos2\phi$ couples $m=+1$ to $m=-1$ directly, so the $|m|=1$ doublet splits at first order in $\delta a$; higher doublets split at higher order or through other Fourier components. Because the operator is real, the split eigenstates can be chosen real, of the form $\cos\phi$ and $\sin\phi$ times real functions, and they carry no current. An anisotropy can split the $\pm m$ pairs of the closed cavity without producing a chiral current.

\section{Discussion}\label{sec:disc}
We end with the optical analogue and the earlier article. The paraxial optical model is examined in Sec.~\ref{sec:opt}. Section~\ref{sec:physical} summarizes the spectral regimes and what can be measured in each, and Sec.~\ref{sec:correction} relates the present results to Ref.~\citep{SilvaOQE2026}.

\subsection{Consistency of the optical model}\label{sec:opt}
The optical part of Ref.~\citep{SilvaOQE2026} used, for a scalar paraxial envelope, the effective operator
\begin{equation}\label{eq:publishedoptic}
H_{\mathrm{opt}}=-\frac{1}{2\kappa}\nabla_\perp^2-\frac{\Om^2r^2}{2\kappa}\partial_\phi^2,
\qquad \kappa=k_0n_0.
\end{equation}
With $[\Om]=\mathrm{length}^{-1}$ the second coefficient in Eq.~\eqref{eq:publishedoptic} has the dimension of length, whereas $\ii\partial_z\psi=H_{\mathrm{opt}}\psi$ requires inverse length. As printed, Eq.~\eqref{eq:publishedoptic} is not dimensionally consistent unless an additional scale is specified.

Apart from this, any rotationally invariant operator of this form conserves the orbital angular momentum of the beam \citep{Allen1992},
\begin{equation}\label{eq:OAMcomm}
[H_{\mathrm{opt}},-\ii\partial_\phi]=0.
\end{equation}
A centered Gaussian contains only $m=0$ and stays in that sector under any rotationally symmetric evolution, including a rotationally symmetric absorbing mask. Throughout such an evolution $\partial_\phi^2\psi=0$, so the twist term of Eq.~\eqref{eq:publishedoptic} acts on it as zero. The large $m\neq0$ output weights reported for a centered Gaussian input therefore cannot come from the stated operator and boundary data, and the optical figures give no support to the spectrum proposed for the closed cylinder. The simulation parameters quoted in Ref.~\citep{SilvaOQE2026} also differ between the text of its Sec.~5 and the captions of its Figs.~6 and 9 (beam waist, window, grid and propagation length), and the twist rate of the run described in its Sec.~5 is not given.

A consistent scalar wave model in curved space starts instead from $(\Delta_g+\kappa^2)\Phi=0$, with the spatial operator of Eq.~\eqref{eq:H}. Even then $[\Delta_g,\partial_\phi]=0$ for constant $\Om$, so axisymmetric coefficients cannot give angular momentum to an axisymmetric input.

Writing $\Phi=\psi\,e^{\ii\kappa z}$ and neglecting $\partial_z^2\psi$, $\partial_\phi\partial_z\psi$ and $\Om^2r^2\partial_z\psi$, which assumes a slowly varying envelope with $|\Om|r\ll1$ across the beam, gives the orbital part of the paraxial equation of Ref.~\citep{SilvaRevisiting2026},
\begin{equation}\label{eq:paraxialcorrect}
\begin{aligned}
\ii\partial_z\psi&=-\frac{1}{2\kappa}\nabla_\perp^2\psi+\frac{\kappa\Om^2r^2}{2}\psi-\Om\hat L_z\psi,\\
\hat L_z&=-\ii\partial_\phi.
\end{aligned}
\end{equation}
Every term has the dimension of inverse length. The right-hand side equals $(2\kappa)^{-1}(-\ii\nabla_\perp-\mathbf A)^2\psi$ with $\mathbf A=\kappa\Om(-y,x)$, which is the Landau problem of Sec.~\ref{sec:travel} with $k$ replaced by $\kappa$; a related Pauli-like equation describes twisted photonic crystal fibers in a frame that follows the twist \citep{Weiss2013}. The same orbital operator, with the vacuum wave number in place of $\kappa$, follows from minimally coupled Maxwell theory in this metric at leading paraxial order, together with a helicity term $-\Om\sigma_3$ that commutes with it; the rigid rotation by $-\Om z$ and the breathing law of Eq.~\eqref{eq:breathing} are derived from the vector equations in Ref.~\citep{SilvaRevisiting2026}. The term $-\Om\hat L_z$, absent from Eq.~\eqref{eq:publishedoptic}, commutes with the rest of the generator and rotates the transverse pattern rigidly by the angle $-\Om z$, while the quadratic term confines the beam in a harmonic well of length $\ell=(\kappa|\Om|)^{-1/2}$. Since the generator commutes with $\hat L_z$, Eq.~\eqref{eq:paraxialcorrect} leaves the conclusion drawn from Eq.~\eqref{eq:OAMcomm} unchanged. With the parameters of Fig.~6 of Ref.~\citep{SilvaOQE2026} ($\lambda_0=1.55\,\mu\mathrm{m}$, $n_0=1.45$, $\Om=5.0\times10^4\,\mathrm{m}^{-1}$, propagation length $0.8\,\mathrm{mm}$), $\ell\simeq1.8\,\mu\mathrm{m}$ is smaller than the input waist $w_0=5\,\mu\mathrm{m}$ and $|\Om|z$ reaches $40$ rad, so neither term is a small correction there. Because $\phi$ is the azimuth of the helicoidal coordinates, the rotation is defined relative to the geometry-adapted frame. Reference~\citep{SilvaRevisiting2026} discusses this frame dependence for the vector problem and shows that a laboratory measurement also requires matching at the interfaces.

\begin{figure*}[t]
\centering\includegraphics[width=\textwidth]{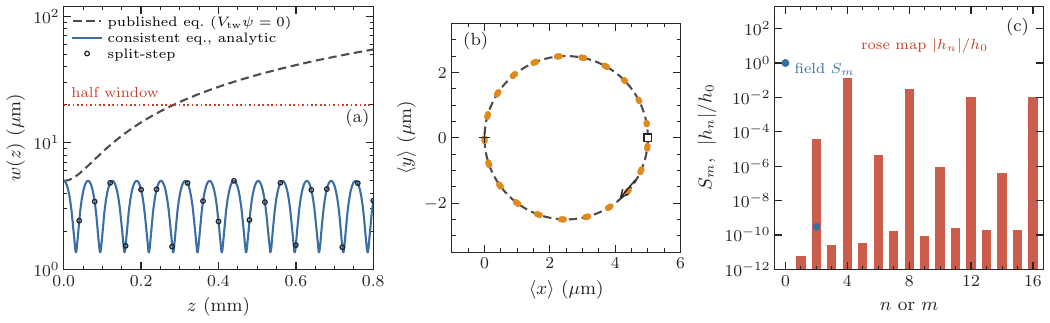}
\caption{Paraxial propagation for the parameters of Fig.~6 of Ref.~\citep{SilvaOQE2026} ($\lambda_0=1.55\,\mu$m, $n_0=1.45$, $\Om=5\times10^4$ m$^{-1}$, $40\,\mu$m window, $600\times600$ grid). (a) Beam radius of a centered Gaussian with $w_0=5\,\mu$m: published equation, which reduces to free diffraction for this input (dashed), Eq.~\eqref{eq:breathing} (solid), and split-step integration of Eq.~\eqref{eq:paraxialcorrect} (circles); the dotted line is the half window. (b) Centroid of an off-axis coherent state ($x_0=5\,\mu$m, width $\sqrt2\ell$) over the first orbit (dots) and the circle $(x_0/2)(1+\cos2\Om z,-\sin2\Om z)$ (dashed); the square marks the start, the cross the axis. (c) For the output field of (a), the orbital-angular-momentum spectrum $S_m$ (dots; $S_0=1-3\times10^{-9}$) and the harmonics $|h_n|/h_0$ of its rose map computed on 720 angular bins of the square window (bars).}\label{fig:opticscorr}
\end{figure*}
The scripts that produced the optical figures of Ref.~\citep{SilvaOQE2026} were re-examined for this work, and their algorithms were re-implemented independently (Appendix~\ref{app:archive}). Figures~6 and 7 there display model fields written down without any propagation: $\psi=e^{-r^2/2\sigma^2}e^{24\ii\phi}$ with $\sigma=4\,\mu\mathrm{m}$, and the ring $\psi=e^{-(r-R_0)^2/2w^2}e^{16\ii\phi}$ with $R_0=15\,\mu\mathrm{m}$ and $w=1.6\,\mu\mathrm{m}$. In each of these figures both panels show the same intensity with different color scales, although the captions describe an input Gaussian and the field after $0.8$ mm. Figure~8 there gives the spectra of synthetic fields, superpositions of angular harmonics with Gaussian weights centered at $m=22$ (bulk) and $m=18$ (edge) and with random phases, so the values of $\langle m\rangle$ shown there were fixed in the input. The rose maps of Fig.~9 there come from free diffraction combined with a rigid rotation of the grid by $\Om\Delta z$ at each step, on a periodic $80\,\mu\mathrm{m}$ window with $800\times800$ points over $5$ mm (the caption gives $500\times500$ points and $10$ mm); the term $V_{\rm tw}$ is not used. None of Figs.~6--9 of Ref.~\citep{SilvaOQE2026} is a solution of its Eqs.~(25)--(27).

\begin{figure}[!htb]
\centering\includegraphics[width=\columnwidth]{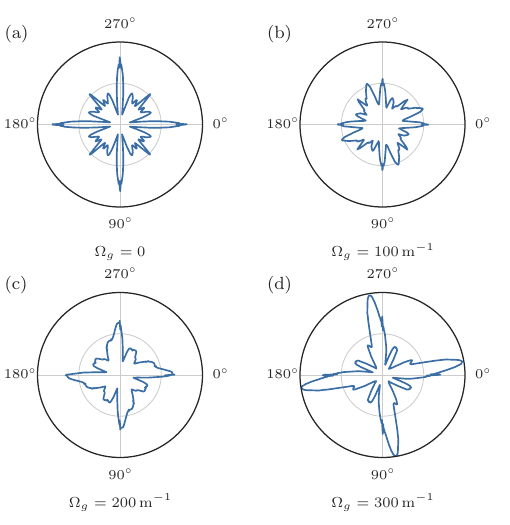}
\caption{Angular mean of the output intensity on 720 bins, computed with the procedure of the script for Fig.~9 of Ref.~\citep{SilvaOQE2026}: free diffraction over $5$ mm with a rigid rotation of the grid by $\Om\Delta z$ at each step $\Delta z=1\,\mu$m, on a periodic $80\,\mu$m window with $800\times800$ points ($w_0=5\,\mu$m, $\lambda_0=1.55\,\mu$m, $n_0=1.45$). All panels are normalized to the largest value of the four. For this axisymmetric input the continuum result is the free beam for every $\Om$, and its angular mean does not depend on $\phi$.}\label{fig:rosesweep}
\end{figure}

The same set contains split-step propagations for the bulk and edge cases whose outputs were not published. They use the grid stated for Fig.~6 of Ref.~\citep{SilvaOQE2026}, an absorbing layer between $17$ and $19\,\mu\mathrm{m}$ from the axis and, for the edge case, an absorbing inner disk of radius $R_{\rm edge}=12\,\mu\mathrm{m}$ with a $1.6\,\mu\mathrm{m}$ taper, a radius not stated in that paper. Their twist step, a forward Euler update with the printed $V_{\rm tw}$, changes the field by less than $10^{-10}$ relative to free diffraction. This is expected for the centered input $\psi=\exp(-r^2/w_0^2)$: for any axisymmetric field $\partial_\phi^2\psi=0$, so the printed $V_{\rm tw}$ of Eq.~\eqref{eq:publishedoptic} vanishes on this input and on its free evolution, and the stated equation reduces to free paraxial diffraction for every value of $\Om$. The Rayleigh range is $z_R=\kappa w_0^2/2=73.5\,\mu\mathrm{m}$ for $w_0=5\,\mu\mathrm{m}$ and $238\,\mu\mathrm{m}$ for $w_0=9\,\mu\mathrm{m}$. After the stated propagation lengths the free beam radius is $54.7\,\mu\mathrm{m}$ for Fig.~6 of Ref.~\citep{SilvaOQE2026} (half window $20\,\mu\mathrm{m}$), $378\,\mu\mathrm{m}$ for the run of its Sec.~5 (half window $125\,\mu\mathrm{m}$), and $681\,\mu\mathrm{m}$ for its Fig.~9 (half window $40\,\mu\mathrm{m}$; $340\,\mu\mathrm{m}$ after the $5$ mm of the script). In the bulk split-step run the output is what the absorbers and the periodic grid leave, and in the edge run the inner disk leaves $3\times10^{-13}$ of the input power. An absorbing disk removes power without reflecting it and cannot create the edge states of a Landau problem, which require a reflecting boundary. The rose-map script has no absorber, so the diffracting beam wraps around the periodic window; at $\Om=0$ the angular mean of the wrapped intensity is a fourfold star at the $10^{-6}$ level, the shape of panel~(a) of Fig.~9 there [Fig.~\ref{fig:rosesweep}(a)]. With $\Om=100$, $200$ and $300\,\mathrm{m}^{-1}$ the script gives the shapes of panels (b)--(d) of that figure [Fig.~\ref{fig:rosesweep}(b)--(d)]. Free diffraction commutes with rotations, so in the continuum this scheme gives the free beam rotated by $\Om z$, which for an axisymmetric input is the free beam itself, and the four maps would coincide. On the grid the maps for $\Om\neq0$ differ from the $\Om=0$ map by $26$, $34$ and $73\%$ (relative rms difference), and by $47$--$65\%$ from the free beam rotated once by the total angle $\Om z$ with the same interpolation. A rotation by an angle that is not a multiple of $\pi/2$ does not map a field that is periodic on the square grid onto another such field, so once the beam has wrapped around the window the interpolated rotation, repeated at each of the $5000$ steps, no longer acts as a rotation of the beam. The sequence of patterns in Fig.~9 of Ref.~\citep{SilvaOQE2026} comes from this procedure. Even for an exactly axisymmetric field an average over the angular bins of a square window depends on angle, because the radial extent of each bin does, and it contains the harmonics $4,8,12,\ldots$ of the square [Fig.~\ref{fig:opticscorr}(c)]. Fourfold and eightfold patterns in such maps do not indicate a change in the angular content of the field. The orbital-angular-momentum spectrum of a centered input stays $S_m=\delta_{m0}$ under any axisymmetric operator and mask, and a mirror-symmetric discretization gives $\langle m\rangle=0$.

With Eq.~\eqref{eq:paraxialcorrect} the same input behaves differently. An axisymmetric Gaussian stays Gaussian and breathes in the harmonic well,
\begin{equation}\label{eq:breathing}
w(z)^2=w_0^2\cos^2(\Om z)+\frac{4\ell^4}{w_0^2}\sin^2(\Om z),
\end{equation}
with period $\pi/|\Om|$. For the parameters of Fig.~6 of Ref.~\citep{SilvaOQE2026}, $w$ oscillates between $5$ and $1.36\,\mu\mathrm{m}$ with period $62.8\,\mu\mathrm{m}$ and stays well inside the window [Fig.~\ref{fig:opticscorr}(a)]. A split-step integration of the full operator, based on the exact factorization $(-\ii\nabla_\perp-\mathbf A)^2=(p_x+\kappa\Om y)^2+(p_y-\kappa\Om x)^2$ with $p_j=-\ii\partial_j$, reproduces Eq.~\eqref{eq:breathing} to better than $5\times10^{-4}w_0$ with step $0.2\,\mu\mathrm{m}$ (and to $1.2\times10^{-2}w_0$ with the published step of $1\,\mu\mathrm{m}$), and it keeps $1-S_0<10^{-8}$. With the masks of the original scripts, Eq.~\eqref{eq:paraxialcorrect} gives $w=3.49\,\mu\mathrm{m}$ at $z=0.8$ mm, as Eq.~\eqref{eq:breathing} does, and the outer absorber removes no power in the bulk case; in the edge case the confined beam lies inside the absorbing disk, and only $10^{-23}$ of the power is left. An off-axis coherent state of width $\sqrt2\,\ell$ centered at $x_0$ keeps its width, and its centroid follows the circle $(x_0/2)(1+\cos2\Om z,\,-\sin2\Om z)$ to $7\times10^{-4}x_0$; this is the cyclotron orbit of the Landau problem [Fig.~\ref{fig:opticscorr}(b)]. For the parameters of Fig.~9 of Ref.~\citep{SilvaOQE2026} and $\Om=100$, $200$ and $300\,\mathrm{m}^{-1}$, $\ell=41.2$, $29.2$ and $23.8\,\mu\mathrm{m}$, the largest beam radius $2\ell^2/w_0$ is $681$, $340$ and $227\,\mu\mathrm{m}$, and the radius after $10$ mm is $573$, $309$ and $32\,\mu\mathrm{m}$, so an $80\,\mu\mathrm{m}$ window is too small for these runs even with the consistent equation. For an axisymmetric Gaussian the diagnostics $\eta_{\rm edge}$ and $\Lambda$ of Ref.~\citep{SilvaOQE2026} follow from $w(z)$, e.g., $\Lambda=\exp[-2(0.8R)^2/w^2]$ without absorber, and carry no information beyond Eq.~\eqref{eq:breathing}.

Describing Maxwell propagation would also need a constitutive tensor \citep{Plebanski1960,LeonhardtPhilbin2006}, a treatment of polarization and a controlled forward-propagation approximation, none of which follows from the quantum Hamiltonian. For this reason we do not reuse the optical figures of Ref.~\citep{SilvaOQE2026}, and we make no claim about photonic chirality.

\subsection{Spectral regimes and observable consequences}\label{sec:physical}
Three spectral situations must be kept apart: a channel with a selected axial momentum $k$, a periodic channel that contains both $\pm k$, and a Dirichlet cavity with standing-wave mixing. The first shows an $m$ asymmetry at fixed $k$, the second is paired by time reversal as a whole, and the third is exactly paired within each $|m|$ sector. Subband mixing, level shifts and current patterns are all compatible with this pairing.

The observable consequences of the cavity model are the finite-radius susceptibility \eqref{eq:chi} and its zero \eqref{eq:criticalgeneral}, the intersubband rule \eqref{eq:D}, the cubic gap \eqref{eq:cubicgap}, and the dipole identities \eqref{eq:TRKz} and \eqref{eq:sumplus}; a hard-wall chiral doublet is not among them. In a tunable system one would measure orbital-resolved energies as $R/L$ varies, probe axial transitions with $\Delta m=0$ and circular transitions with $\Delta m=\pm1$, and check that the spectral sum follows $1+\Om^2\langle r^2\rangle$. In a semiconductor the metric effective-mass model would first have to be derived from strain, band structure and confinement, and spin--orbit coupling, Zeeman fields, contact self-energies and finite barriers would add terms that can change these conclusions.

The anticrossing within one $m$ sector and the associated transfer of dipole strength (Secs.~\ref{sec:anticross} and \ref{sec:spectroscopy}) never split a time-reversal pair. We have not treated scattering with open leads, thermal occupation, a microscopic realization, a realistic electromagnetic model, or electrical-network implementations.

\subsection{Relation to the 2026 Optical and Quantum Electronics article}\label{sec:correction}
The statements, equations and figures of Ref.~\citep{SilvaOQE2026} are listed in Table~\ref{tab:status} (Appendix~\ref{app:status}) together with their status in the present analysis. Each quantitative statement that has a counterpart in the corrected model is re-derived in the section cited in the table. The table covers the operator-domain error and the separate dimensional and normalization problems; how the published optical figures were produced is described in Sec.~\ref{sec:opt}. Section~VII\,A of Ref.~\citep{SilvaRevisiting2026} had already pointed out that the hard-wall version of Ref.~\citep{SilvaOQE2026} needs a separate treatment of the boundary, that $E(m)=E(-m)$ for the finite cylinder, and that its optical part must be reanalyzed. The present paper provides that treatment. One of us (E.O.S.), the sole author of Ref.~\citep{SilvaOQE2026}, has requested its retraction from that journal.

\section{Conclusions}\label{sec:concl}
We have treated the scalar Hamiltonian of the helicoidal metric with the axial Dirichlet domain included in the operator. The axial derivative couples real standing-wave subbands and cannot be replaced by a fixed axial wave number. Time reversal and rotational symmetry together exclude the splitting between $+m$ and $-m$ proposed earlier for the closed cavity. Traveling waves on an unbounded or periodic axis keep their fixed-$k$ description, which is a Landau problem with effective field proportional to $\Om k$; there the $m$ asymmetry is the paramagnetic term.

With finite radial confinement the leading twist correction changes sign at a critical aspect ratio, while for an infinite radius the response is linear in $|\Om|$. A gauge-adapted variational form and the original basis give consistent spectra and currents. In any stationary eigenstate the total axial flux through each cross section vanishes, although local circulation is allowed. Longitudinal and transverse dipole sum rules test the full intersubband spectrum.

The main predictions are an isolated avoided crossing within one $m$ sector whose gap grows as the cube of the twist when the aspect ratio is retuned, and a circular-dipole line whose strength starts at second order in the twist. Neither effect breaks the pairing of time-reversed sectors. We have also given the inverse-mass tensor that a single-band scalar realization would need. Whether a rolled semiconductor provides it is an open question, and finite barriers, spin--orbit coupling and realistic linewidths would add terms that must be justified separately. The angular-momentum generation from a centered input reported for the published paraxial model does not follow from its own equations, and the original scripts show that the optical figures were not obtained by solving them. The consistent paraxial equation, Eq.~\eqref{eq:paraxialcorrect}, is a Landau problem with an additional rigid rotation and also conserves $\hat L_z$. Every quantitative statement of Ref.~\citep{SilvaOQE2026} with a counterpart in the corrected model is re-derived here (Appendix~\ref{app:status}): the effective-potential minimum and the energy scale survive for traveling states, while the current profile, the optical patterns and the estimates for the proposed systems change. The corrected scalar boundary-value theory makes predictions that can be checked directly; a microscopic or vector-Maxwell description of an actual device is still missing.

\appendix
\section{Analytic identities and the gauge-adapted matrix}\label{app:analytic}
Here we collect the analytic results used in the main text. They are the Bessel normalization and radial moment, the subband sum with its dimensional checks, the derivation of the gauge-adapted matrix, and the proof of the variational bound of Sec.~\ref{sec:crossover}.

\subsection{Bessel normalization and radial moment}\label{app:bessel}
The identity
\begin{equation}
\int_0^1 x\,\dd x\,J_\ell(j_{\ell s}x)^2
=\tfrac12 J_{\ell+1}(j_{\ell s})^2
\end{equation}
normalizes Eq.~\eqref{eq:Bessel}. Repeated use of the Bessel differential equation, integration by parts, and $J_\ell(j_{\ell s})=0$ yields
\begin{equation}
\frac{\int_0^1x^3J_\ell(j_{\ell s}x)^2\,\dd x}{\int_0^1xJ_\ell(j_{\ell s}x)^2\,\dd x}
=\frac13\left[1+\frac{2(\ell^2-1)}{j_{\ell s}^2}\right].
\end{equation}
Gauss--Legendre quadrature of this ratio for several $\ell$ and $s$ agrees with the closed form to machine precision.

Current densities depend on the normalization convention, which we fix as follows. If $F(r)$ is normalized with $\int_0^\infty r\,\dd r\,|F|^2=1$, the corresponding wavefunction $F(r)e^{\ii m\phi}/\sqrt{2\pi}$ has unit norm in two dimensions. For $n_r=0$ and $\ell=|m|$, the normalized full transverse wavefunction may be written
\begin{equation}
\begin{aligned}
\psi_{0m}(r,\phi)
&=\frac{e^{\ii m\phi}}{\sqrt{\pi\ell_{\rm ho}^2\Gamma(\ell+1)}}\\
&\quad\times\left(\frac{r}{\ell_{\rm ho}}\right)^\ell
 e^{-r^2/(2\ell_{\rm ho}^2)},\\
\ell_{\rm ho}&=\beta^{-1/2}.
\end{aligned}
\end{equation}
The normalization coefficient in Eq.~\eqref{eq:laguerre} follows from
\begin{equation}\label{eq:Laguerre_orth}
\int_0^\infty e^{-x}x^{|m|}
 \bigl[L_{n_r}^{|m|}(x)\bigr]^2\dd x
=\frac{\Gamma(n_r+|m|+1)}{n_r!}.
\end{equation}
Equivalently, with $u(r)=\sqrt r\,F(r)$, one has
\begin{equation}\label{eq:Laguerre_norm}
\int_0^\infty |u(r)|^2\dd r
=\int_0^\infty r|F(r)|^2\dd r=1.
\end{equation}
With the coefficient printed in Eq.~(15) of
Ref.~\citep{SilvaOQE2026}, where $n_r!$ appears in the
denominator, the squared norm would be $1/(n_r!)^2$. The difference
appears only for $n_r\ge2$, since both factorials equal one for
$n_r=0,1$. The published applications used the radial ground state
$n_r=0$, which this error leaves unchanged; the general formula
must still be corrected. The axial-domain problem is separate and affects the
cavity interpretation already at $n_r=0$.

The transformation $u(r)=\sqrt r\,F(r)$ leaves the coefficient $\mathcal N_{n_r,m}$ unchanged while replacing the radial measure $r\,\dd r$ by $\dd r$. Both this distinction and the azimuthal factor $1/\sqrt{2\pi}$ matter in current densities. Without an additional angular factor, the state of Eq.~(20) of Ref.~\citep{SilvaOQE2026} has norm $2$ in the plane, i.e., it differs by a factor $\sqrt2$ from a unit-normalized state.

\subsection{Subband sum and dimensional checks}\label{app:sum}
With $\chi_n$ of Eq.~\eqref{eq:axbasis}, $\langle p|p_z|n\rangle=-\ii\hbar D_{pn}/L$. Equation~\eqref{eq:sump} is equivalent to the dimensionless series
\begin{equation}
\sum_{p\ne n}\frac{D_{pn}^2}{\pi^2(n^2-p^2)}=-\frac14.
\end{equation}
Numerical summation of the first 3000 terms for $n=1,\ldots,4$ verifies the stated constant with an error below $5\times10^{-10}$. Dimensional analysis of the axial/traveling cross term gives $[\hbar^2\Om mk/M]=\mathrm{energy}$, while $[\hbar\Om mk]=\mathrm{energy}/\mathrm{length}$ and thus cannot be the complete energy coefficient in standard SI units. Likewise, an optical forward generator $\ii\partial_z$ has units $\mathrm{length}^{-1}$; the prefactor of Eq.~\eqref{eq:publishedoptic} has units of length if $\Om$ retains the dimensions in Eq.~\eqref{eq:metric}.

\subsection{Derivation of the gauge-adapted variational matrix}\label{app:gaugematrix}
The equivalence of Eqs.~\eqref{eq:positive} and \eqref{eq:gaugeform} follows from applying the unitary map $u_m=e^{\ii\alpha_m z}v_m$ to the quadratic form; no manipulation of $p_z$ on the Dirichlet domain is involved. For the dimensionless basis
$V=\sum_{s,n}c_{sn}\mathcal B_{|m|s}(x)s_n(\zeta)$,
where $s_n(\zeta)=\sqrt2\sin(n\pi\zeta)$, define the axial matrices
\begin{equation}
\begin{aligned}
Z_{pn}&=\int_0^1s_p(\zeta)(\zeta-\tfrac12)
 s_n(\zeta)\,\dd\zeta,\\
Z^{(2)}_{pn}&=\int_0^1s_p(\zeta)(\zeta-\tfrac12)^2
 s_n(\zeta)\,\dd\zeta.
\end{aligned}
\end{equation}
Let $\mathcal B_s=\mathcal B_{|m|s}$, $q=q_m$ and primes denote
$\partial_x$. Define the radial symmetric matrices
\begin{align}
K_{st}&=\int_0^1x\,\dd x\left[\mathcal B_s'\mathcal B_t'+\frac{m^2a^2}{x^2(a^2+b^2x^2)}\mathcal B_s\mathcal B_t\right],\\
S_{st}&=\int_0^1x\,\dd x\,(a^2+b^2x^2)\mathcal B_s\mathcal B_t,\nonumber\\
T_{st}&=\int_0^1x\,\dd x\,(q')^2\mathcal B_s\mathcal B_t,\nonumber
\end{align}
and the real antisymmetric matrix
\begin{equation}
A_{st}=\int_0^1x\,\dd x\,q'(\mathcal B_s'\mathcal B_t-\mathcal B_s\mathcal B_t'),\qquad A_{st}=-A_{ts}.
\end{equation}
The gauge-adapted Galerkin matrix is then
\begin{equation}\label{eq:gaugematrix}
\begin{aligned}
\frac{H_m^{\rm(gauge)}}{E_R}
&=(K\otimes I_z)
 +(S\otimes\mathrm{diag}[(\pi n)^2])\\
&\quad+(T\otimes Z^{(2)})+\ii(A\otimes Z).
\end{aligned}
\end{equation}
The final term is Hermitian because $A$ is antisymmetric and $Z$ symmetric. At $b=0$, $q'=0$, $K_{st}=j_{|m|s}^2\delta_{st}$, and Eq.~\eqref{eq:gaugematrix} returns the exact Bessel--sine spectrum. The two bases span different finite-dimensional variational subspaces and have different convergence rates, but represent the same self-adjoint continuum Hamiltonian.

\subsection{Proof of the finite-wall variational bound}\label{app:cutoff}
Write $F_{\infty}(r)=\sqrt{2\beta}\,e^{-\beta r^2/2}$ for the normalized radial ground state, where $\beta=|\Om|k_n$ and $\int_0^\infty r\,\dd r\,|F_\infty|^2=1$. Let $\chi_R(r)$ be unity for $r\le R/2$, decrease linearly to zero on $[R/2,R]$, and vanish for $r\ge R$. Domain monotonicity gives $E^{(R)}_{0,0,n}\ge E^{(\infty)}_{0,0,n}$, and the ground-state representation of the form gives
\begin{equation}
\frac{Q[\chi_RF_\infty]}{\|\chi_RF_\infty\|^2}-E^{(\infty)}_{0,0,n}
=\frac{\hbar^2}{2M}\frac{\int_0^Rr\,\dd r\,|\chi_R'|^2|F_\infty|^2}{\int_0^Rr\,\dd r\,\chi_R^2|F_\infty|^2}.
\end{equation}
The cutoff derivative satisfies $|\chi_R'|\le2/R$. The Gaussian tail beyond the halfway radius equals $e^{-\beta R^2/4}$. Hence the numerator is no larger than $(4/R^2)e^{-\xi_n/4}$, and the denominator is at least $1-e^{-\xi_n/4}$. The min--max principle now yields Eq.~\eqref{eq:cutoffbound}.

\section{Numerical methods and dipole checks}\label{app:numerics}
The numerical methods are described below, together with the checks behind the dipole results of Sec.~\ref{sec:spectro}.

\subsection{Numerical implementation and verification}\label{app:numerical}
The dimensionless Galerkin matrix is assembled as
\begin{equation}
\begin{aligned}
\frac{H_m}{E_R}&=\mathrm{diag}(j_{\ell s}^2)\otimes I_z
+I_r\otimes\mathrm{diag}[(\pi a n)^2]\\
&\quad+b^2X\otimes\mathrm{diag}[(\pi n)^2]
+2\ii mb\,I_r\otimes D.
\end{aligned}
\end{equation}
The radial quadratures and $D$ are tested for orthogonality, symmetry/antisymmetry, and Hermiticity. The $m$ and $-m$ matrices are complex conjugates. In the finite-difference implementation, the transformation $v=\sqrt r\,u$ yields the radial expression $-\partial_r^2+(m^2-1/4)/r^2$. We discretize $\partial_z$ by an antisymmetric centered first difference and $-\partial_z^2$ by a symmetric centered second difference, with both endpoints eliminated by Dirichlet data. The finite-difference check is restricted to $|m|\ge1$ because of the critical inverse-square behavior at the axis for $m=0$.

Truncations that converge the energies need not give the same accuracy for local currents. The current maps are computed in the gauge-adapted representation and checked with the flux identity (Table~\ref{tab:currentaudit}) and the pointwise comparison of Sec.~\ref{sec:currentrefinement}. The $z$-dipole sum rule was evaluated with the complete eigensystem of a $14\times30$ basis, summing all transitions from the ground state; the truncation misses a positive high-energy tail, which explains the small deficit with respect to the commutator value.

\subsection{Cubic two-level reduction and axial parity}\label{app:cubic}
The exact gauge-adapted matrix is Eq.~\eqref{eq:gaugematrix}. For fixed $a$ and small $b$, its radial matrices have expansions
\begin{align}
 K_{st}&=j_{|m|s}^2\delta_{st}-\frac{m^2b^2}{a^2}\delta_{st}+O(b^4),\\
 S_{st}&=a^2\delta_{st}+b^2X_{st},\\
 A_{st}&=-\frac{2mb^3}{a^4}
 \int_0^1x^2(\mathcal B_s'\mathcal B_t-\mathcal B_s\mathcal B_t')\,\dd x+O(b^5),\\
 T_{st}&=O(b^6).
\end{align}
These expansions show why the states $(1,2)$ and $(2,1)$, which cross at $b=0$, cannot couple at order $b^2$: $S_{st}$ conserves the axial label, while $A_{st}Z_{pn}$ changes radial and axial labels simultaneously. The first nonzero off-diagonal term of the projected Hamiltonian is cubic. We evaluated its coefficient by Gauss--Legendre quadrature and compared it with the small-$b$ behavior of the full optimized Galerkin gap. The terms of order $b^2$ shift the two diagonal energies by $b^2\pi^2n^2 C_{1,s}-m^2b^2/a^2$. Because $C_{1,s}=1/3$, their difference produces $a_\ast^2=a_0^2-b^2/3+O(b^4)$, whereas the common $-m^2b^2/a^2$ cancels. The effective two-level gap is $2|C_{AB}||b|^3+O(|b|^5)$ only after this diagonal detuning has been minimized. The reduction holds for small $b$ and isolated neighboring levels; it says nothing about the high-energy spectrum or multiple simultaneous degeneracies.

\subsection{Circular oscillator sum and dipole matrix in the original basis}\label{app:circsum}
For the full scalar metric Hamiltonian and an $m=0$ eigenstate $|i\rangle$, define $D_+=x+\ii y$ and $D_-=x-\ii y$. Completeness, $g^{xx}=g^{yy}=1$ and $g^{xy}=0$ imply
\begin{equation}
 \sum_f(E_f-E_i)\bigl(|\langle f|x|i\rangle|^2+|\langle f|y|i\rangle|^2\bigr)=\frac{\hbar^2}{M}.
\end{equation}
Rotational symmetry partitions the two transverse circular channels into $m_f=\pm1$. For the nondegenerate $m_i=0$ ground state, time reversal makes the two energy-weighted sums equal. With the unnormalized $D_\pm$ convention, each sum is $\hbar^2/M$, yielding Eq.~\eqref{eq:sumplus}. In the original untransformed Bessel--sine basis, the circular dipole matrix between adjacent $m$ sectors is especially simple:
\begin{equation}\label{eq:plainmatrix}
\begin{aligned}
&\frac{\langle s',p,m+1|D_+|s,n,m\rangle}{R}\\
&\quad=\delta_{pn}\int_0^1x^2
\mathcal B_{|m+1|,s'}(x)\mathcal B_{|m|,s}(x)\,\dd x.
\end{aligned}
\end{equation}
This expression does not involve the gauge-adapted form. We used it for all transition weights and for the truncated sum $S_+$, and to cross-check the gauge-adapted amplitudes of Eq.~\eqref{eq:relativephase}, which were evaluated by quadrature. The activated strength at $b=2$ is $0.1063709848$ in the original $(14,22)$ basis and $0.1063708147$ in the gauge-adapted one, a difference consistent with their truncation errors. For $(N_r,N_z)=(16,24)$ the original-basis sum is $S_+=1.999985672$ at $b=2$, a deficit of $1.43\times10^{-5}$ from the exact value. The sum is sensitive to any missing high-lying radial transitions.

\subsection{Perturbative strength and additional checks}\label{app:activationaudit}
Let $\mathcal B_{\ell s}(x)=R B_{\ell s}(Rx)$ with $x=r/R$ and
$\int_0^1 x\,\dd x\,\mathcal B_{\ell s}^2=1$. Consider the
$|m_i=0,s=1,n=1\rangle$ ground state and the $m_f=1$ eigenstate
which is connected continuously to $|m_f=1,s=1,n=2\rangle$ at zero twist.
At $b=0$ the circular dipole conserves the axial index and this
transition vanishes. The linear perturbation $2\ii m bD_{pn}$ in
Eq.~\eqref{eq:matrix} acts only in the axial subspace. Consequently,
only the $n=1$ admixture in the final state contributes to the
first-order transition amplitude. Nondegenerate perturbation
theory gives
\begin{equation}\label{eq:activation_amp}
\begin{aligned}
\frac{\langle f|d_+|i\rangle}{R}
&=-\frac{2\ii b D_{12}}{3\pi^2a^2}\mathcal I_{10}
+O(b^3),\\
D_{12}&=-\frac83,\\
\mathcal I_{10}&=\int_0^1 x^2\mathcal B_{1,1}(x)
\mathcal B_{0,1}(x)\,\dd x.
\end{aligned}
\end{equation}
The overall phase in Eq.~\eqref{eq:activation_amp} depends on the
basis convention, but its magnitude does not. The unperturbed
transition gap, in units $E_R$, is
\begin{equation}\label{eq:delta12}
\Delta_{12}(a)=j_{1,1}^2-j_{0,1}^2+3\pi^2a^2.
\end{equation}
Hence the leading strength is
\begin{equation}\label{eq:activatedcoeff}
\mathcal A(a)=\Delta_{12}(a)
\left(\frac{2D_{12}\mathcal I_{10}}{3\pi^2a^2}\right)^2.
\end{equation}
Gauss--Legendre quadrature gives
$\mathcal I_{10}=0.4654514086\ldots$ and
$\mathcal A(2)=0.0559408010\ldots$.
The linear transition amplitude is pure imaginary in the original
real Bessel--sine basis. Since the opposite-twist Hamiltonians are
related by complex conjugation at fixed $m$, the squared transition
strength is even in $b$; hence its next correction is $O(b^4)$.

\begin{table}[t]
\centering\small
\caption{Strength of the activated circular transition at $a=2$
in the untransformed Bessel--sine basis, $(N_r,N_z)=(14,20)$. The last
column approaches $\mathcal A(2)=0.0559408010\ldots$ as $b\to0$.
The unnormalized dipole $d_+=x+\ii y$ is used.}
\label{tab:independ_activation}
\begin{tabular}{rrr}\toprule
$b$ & $f_+^{\rm activated}$ & $f_+/b^2$\\\midrule
$0.025$ & $0.0000349580$ & $0.05593276$\\
$0.050$ & $0.0001397716$ & $0.05590865$\\
$0.100$ & $0.0005581232$ & $0.05581232$\\
$0.150$ & $0.0012521748$ & $0.05565221$\\
$0.300$ & $0.0049317405$ & $0.05479712$\\\bottomrule
\end{tabular}
\end{table}

Direct diagonalization in the untransformed basis gives the optimized
gaps
$\Delta_{\rm gap}/E_R=0.00284790328$ at $b=0.12$ and
$0.04438638433$ at $b=0.30$ in a $(10,16)$ basis. Their ratios to
$1.6487752187\,b^3$ are $0.999586$ and $0.997067$.
For the transverse sum at $a=b=2$, a $(14,20)$ basis yields
$S_+=1.99997918$, approaching the exact value $2$ in
Eq.~\eqref{eq:sumplus}. The matrices satisfy
$H_{-m}=H_m^*$ exactly in this basis, a direct check of time-reversal
symmetry.

The longitudinal sum \eqref{eq:TRKz} was evaluated in the same
basis with the axial coordinate matrix \eqref{eq:zselection}. For $a=b=2$, the pairs
$(S_z,1+\Om^2\langle r^2\rangle)$ in a $(14,26)$ basis are
$(1.14622537,1.14628829)$ for $m=0$ and
$(1.26368838,1.26375705)$ for $m=1$.
The small deficits come from the finite spectral cutoff. These checks
test the main spectroscopic results at the reported parameters but do
not give error bounds for every plotted quantity.

A third, separately written implementation of the original matrix
gives $2|C_{AB}|=1.6487752187433$ and
$\mathcal A(2)=0.0559408010177$ by direct quadrature. With
$(N_r,N_z)=(12,16)$ at $a=b=2$ it gives an activated strength
$0.1063711740$ and a truncated $m_f=+1$ sum $1.9999681477$. At
$b=0.025$ and $0.05$ the ratios $f_+/b^2$ are $0.0559327604$ and
$0.0559086469$, approaching the perturbative coefficient as $b\to0$.
The residual of $H_{-m}=H_m^*$ is zero to floating-point accuracy.

\section{Status of the statements of the 2026 Optical and Quantum Electronics article}\label{app:status}
For each statement, equation and figure of Ref.~\citep{SilvaOQE2026}, Table~\ref{tab:status} gives its status in the present analysis and the section of this paper where it is re-derived or discussed.
\begin{table*}[t]
\centering\small
\caption{Scientific status of statements in Ref.~\citep{SilvaOQE2026} within the specified scalar model.}
\label{tab:status}
\begin{tabular}{p{0.31\textwidth}p{0.60\textwidth}}\toprule
Earlier statement & Present conclusion\\\midrule
Helicoidal metric and Laplace--Beltrami operator & Correct for constant $\Om$; a well-defined nonflat Riemannian background.\\
Fig.~1: transverse planes rotated by $\Om\Delta z$ & That picture describes flat space in helicoidal coordinates. In Eq.~\eqref{eq:metric} the curves orthogonal to $\partial_z$ obey $\dd z=-\Om r^2\dd\phi$, Fig.~\ref{fig:geometry}.\\
Traveling-wave Landau-like oscillator & Valid for a conserved axial wave number $k\ne0$, with the energy dimensions corrected.\\
Fig.~2: $r_0=74.3$ nm, $V_{\min}=1.564$ meV, $\omega=7.902\times10^{11}$ s$^{-1}$ & Correct for the reduced function $u$ of a traveling state with $k=\pi/L$, Eq.~\eqref{eq:Veffu}; with CODATA constants $V_{\min}=1.565$ meV and $\omega=7.906\times10^{11}$ s$^{-1}$; the printed values follow from the rounded constants $\hbar=1.054\times10^{-34}$ J\,s and $m_e=9.11\times10^{-31}$ kg of the original script. Not a level of the cavity.\\
$k=n\pi/L$ and $\partial_z=\pm\ii k$ under Dirichlet walls & Invalid for a closed finite axial box. The derivative is off-diagonal, Eq.~\eqref{eq:D}.\\
Linear $m\Om n/L$ splitting of closed cavity levels & Excluded exactly by spinless time reversal for a rotationally symmetric Dirichlet domain.\\
$2\hbar\omega\simeq1.04$ meV as a $\pm m$ split & Not a $\pm m$ splitting of the cavity. The same energy is the $m=0$ radial spacing for $R=\infty$, Eq.~\eqref{eq:spacing}, i.e., the cyclotron energy $\hbar\omega_c$ of the traveling problem.\\
Geometry-selected equilibrium charge current & Not implied by the scalar Hamiltonian; prepared $\pm m$ partners carry opposite current.\\
Radial equation and normalization in Eqs.~(9)--(15) & The $-1/4$ centrifugal term belongs to the reduced radial function $u=\sqrt r\,F$, not the radial factor $F$ in Eq.~(9). The radial norm must use the correct measure; the printed $n_r!$ denominator in Eq.~(15) must be replaced by a numerator. The difference starts only at $n_r\ge2$ and does not change the reported $n_r=0$ special-case normalization.\\
Printed last term of Eq.~(16) & In SI units the traveling-wave term is $-\hbar^2\Om mk/M$; the printed $-\hbar\Om m n\pi/L$ has the wrong dimensions. Neither expression yields a $\pm m$ split in the Dirichlet cavity.\\
Radial state of Eq.~(17) with $e^{\ii m\phi}$ & For $m\neq0$ the $m=0$ Gaussian multiplied by $e^{\ii m\phi}$ is not an eigenstate, and $|\Psi|^2/r$ diverges on the axis; the $n_r=0$ states are those of Eq.~\eqref{eq:laguerre}.\\
Azimuthal current of Eq.~(21) & The current in the curved metric contains the term from $g^{\phi z}\partial_z$ in addition to the flat-space $m/r$ term. With it, the integrated circulation of every traveling state has the sign of $-\Om k$, not of $m$, Eq.~\eqref{eq:Lkin}. Partners carry opposite currents, and the metric alone selects no equilibrium current.\\
Current maximum at $\ell_{\rm ho}/\sqrt2=56.4$ nm (Figs.~3--5) & With the metric term the density peaks at $\ell_{\rm ho}=79.8$ nm, where the local current changes sign; extrema at $37.4$ and $120.5$ nm, Eq.~\eqref{eq:Jtravel} and Fig.~\ref{fig:travelprofiles}. The original script for Figs.~4 and 5 also builds the density from $u=\sqrt r\,F$ instead of $F$, which moves the $m=1$ ring to $\sqrt{3/2}\,\ell_{\rm ho}\simeq98$ nm.\\
Figs.~6, 7, 9: bulk, edge and rose patterns; $\eta_{\rm edge}$, $\Lambda$ & Figs.~6 and 7 show model fields with the phases $e^{24\ii\phi}$ and $e^{16\ii\phi}$, written down without propagation. Fig.~9 combines free diffraction with a rigid rotation on a periodic window; its four panels are reproduced in Fig.~\ref{fig:rosesweep} and result from the beam wrapping around that window and from the interpolated rotation of the grid. For a centered Gaussian the printed $V_{\rm tw}$ vanishes. With Eq.~\eqref{eq:paraxialcorrect} the beam breathes, Eq.~\eqref{eq:breathing}, and the absorbing inner disk of the original script ($R_{\rm edge}=12\,\mu$m) removes the confined beam instead of producing an edge state.\\
Optical OAM redistribution from a centered Gaussian (Fig.~8) & Not possible with the printed rotationally invariant operator, nor with the consistent paraxial equation \eqref{eq:paraxialcorrect}, which also conserves $\hat L_z$; $S_m=\delta_{m0}$, Fig.~\ref{fig:opticscorr}(c). The published spectra are those of synthetic fields with Gaussian weights centered at $m=22$ and $m=18$.\\
Tolerances: $\Om(z)$ and ellipticity (Secs.~2 and 7.1) & $\Om(z)$: Sec.~\ref{sec:general}, where an axial average is not sufficient. A $\cos2\phi$ anisotropy splits the $|m|=1$ pair at first order into real standing waves without current.\\
Rashba comparison (Sec.~7.2) & $2\alpha_Rk\approx0.075$--$0.19$ meV, comparable to or larger than the anticrossing gaps of Table~\ref{tab:gaps}; there is no $1.04$ meV doublet to compare with.\\
Proposed systems and detection (Sec.~7, Table~1) & Restated in Table~\ref{tab:platforms} and Sec.~\ref{sec:implementation}: no $\pm m$ doublet and no equilibrium current; cold-atom cavity scales below $1$ Hz.\\\bottomrule
\end{tabular}
\end{table*}

\section{Scope, conventions, and reproducibility}\label{app:scoperepro}
We state here the scope of the model and the unit conventions, and describe how the calculations were done and checked.

\subsection{Experimental-model scope and dimensional conventions}\label{app:scope}
For any displayed energy $\Delta E$ in meV, the conversion to frequency is $\nu=\Delta E/h$, or $1\,\mathrm{meV}/h\simeq0.241799\,\mathrm{THz}$. The radial unit $E_R=\hbar^2/(2MR^2)$ remains constant when the figures scan $a$ with $R$ and $M$ fixed. Fixing $b$ while varying $a$ changes the physical $L=R/a$ and necessarily changes the physical twist according to $\Om=b/(aR)$; an experiment with fixed $\Om$ follows a different path $b=\Om R\,a$. Absolute electronic absorption would require the state charge, dielectric environment, external field coupling, occupation factors, spin-orbit corrections and linewidths. None of these enters the present scalar model.

\subsection{Computational procedures and reproducibility}\label{app:archive}
The calculations used separately assembled Bessel--sine and gauge-adapted Galerkin matrices, finite-difference checks, and direct evaluations of the axial and circular dipole sum rules. The small-twist formulas were checked against analytic Bessel identities and direct eigenvalue minimization. Computed currents were compared on an interior grid against larger-basis reference states, and their integrated axial flux was tested against the exact conservation argument. The nonuniform-twist matrices were constructed in Hermitian form with the derivative-ordering terms retained. The scripts used for the figures of Ref.~\citep{SilvaOQE2026} were read but not executed; their algorithms (effective potential, model fields, split-step propagation with masks, synthetic spectra and rose maps) were re-implemented and run with the parameters found in them, and the outputs were compared with the published and unpublished images. The displayed dimensionless parameters, boundary conditions, basis conventions and numerical truncations are specified in the text, including Tables~\ref{tab:localcurrent} and \ref{tab:gaps}. The scripts for all calculations, figures and verification tests are provided as ancillary files (see Data availability). No experimental measurements were used.

\section*{Declaration of the use of AI tools}
Two large language models were used in this work. GPT-6 Pro (OpenAI) was used to review the Python scripts, to check the analytical and numerical calculations, and to revise the English of the manuscript. Claude Opus 5.5 (Anthropic) was used to write the Python scripts for the numerical calculations, the verification tests and the figures, including the independent re-implementation of the routines behind the optical figures of Ref.~\citep{SilvaOQE2026} (Sec.~\ref{sec:opt}); to review the mathematical calculations; and to draft and edit parts of the text. The numerical results were checked with reproducible test scripts, which are available as stated below. Neither tool is listed as an author. The authors reviewed and edited all output of these tools and take full responsibility for the content of this article.

\section*{Data availability}
The Python scripts that produce the numerical results, tables and figures of this article, the independent re-implementations of the routines behind the optical figures of Ref.~\citep{SilvaOQE2026}, and the verification tests with their output logs are provided as ancillary files with the arXiv version of this article. No experimental data were used.

\begin{acknowledgments}
This work was supported by Conselho Nacional de Desenvolvimento Cient\'ifico e Tecnol\'ogico (CNPq) (grant 305427/2026-1), Funda\c c\~ao de Amparo \`a Pesquisa e ao Desenvolvimento Cient\'ifico e Tecnol\'ogico do Maranh\~ao (FAPEMA) (grant UNIVERSAL-06395/22), and Coordena\c c\~ao de Aperfei\c coamento de Pessoal de N\'ivel Superior (CAPES) -- Brazil (Code 001).
\end{acknowledgments}

\bibliographystyle{apsrev4-2}
\bibliography{references}
\end{document}